\documentclass[11pt]{article}

\usepackage{amsmath,amsfonts,amssymb,bm}

\usepackage{cancel}
\usepackage{tikz-cd}
\usepackage{float}
\usepackage{diagbox,pdflscape}

\usepackage[numbers,sort&compress]{natbib}
\usepackage{graphicx}

\usepackage{hyperref}
\hypersetup{
   colorlinks   =  true,
    linkcolor    = cyan,
    citecolor    = red,
     urlcolor	=magenta,     
}

\usepackage{amsthm}
\theoremstyle{definition} 

\theoremstyle{theorem} 

\newcommand\be{\begin{equation}}
\newcommand\ee{\end{equation}}
\newcommand\bea{\begin{eqnarray}}
\newcommand\eea{\end{eqnarray}}

\def\dd{{\rm d}}
\newcommand{\bq}{\mathbf{q}}
\newcommand{\bp}{\mathbf{p}}
\def\la{{\lambda}}
\def\om{{\omega}}
\def\RR{\mathbb{R}}
\def\te{\theta}

\newcommand{\hbq}{\hat{\mathbf{q}}}
\newcommand{\hbp}{\hat{\mathbf{p}}}
\newcommand{\hH}{\hat{\cal{H}}}
\newcommand{\hC}{\hat{C}}
\newcommand{\hI}{\hat{I}}

\newcommand{\red}[1]{{\color{red}#1}}

\newcommand{\hq}{\hat{q}}
\newcommand{\hp}{\hat{p}}

\title{Shannon information entropy for a quantum nonlinear oscillator on a space of non-constant curvature}

\begin{document}

\maketitle

\begin{center}

{\sc Angel Ballesteros$^{1}$ and Ivan Gutierrez-Sagredo$^{2}$}

\medskip
{$^1$Departamento de F\'isica, Universidad de Burgos, 
09001 Burgos, Spain}

{$^2$Departamento de Matem\'aticas y Computaci\'on, Universidad de Burgos, 
09001 Burgos, Spain}
\medskip
 
e-mail: {\href{mailto:angelb@ubu.es}{angelb@ubu.es}, \href{mailto:igsagredo@ubu.es}{igsagredo@ubu.es}}

\end{center}

\begin{abstract}

The so-called Darboux III oscillator is an exactly solvable $N$-dimensional nonlinear oscillator defined on a  radially symmetric space with non-constant negative curvature. This oscillator can be interpreted as a smooth  (super)integrable deformation of the usual $N$-dimensional harmonic oscillator in terms of a non-negative parameter $\lambda$ which is directly related to the curvature of the underlying space. In this paper, a detailed study of the Shannon information entropy for the quantum version of the Darboux III oscillator is presented, and the interplay between entropy and curvature is analysed. In particular, analytical results for the Shannon entropy in the position space can be found in the $N$-dimensional case, and the known results for the quantum states of the $N$-dimensional harmonic oscillator are recovered in the limit of vanishing curvature $\lambda \to 0$. However, the Fourier transform of the Darboux III wave functions cannot be computed in exact form, thus preventing the analytical study of the information entropy in momentum space. Nevertheless, we have computed the latter numerically both in the one and three-dimensional cases and we have found that by increasing the absolute value of the negative curvature (through a larger $\lambda$ parameter) the information entropy in position space increases, while in momentum space it becomes smaller. This result is indeed consistent  with the spreading properties of the wave functions of this quantum nonlinear oscillator, which are explicitly shown. The sum of the entropies in position and momentum spaces has been also analysed in terms of the curvature: for all excited states such total entropy decreases with $\lambda$, but for the ground state the total entropy is minimised when $\lambda$ vanishes, and the corresponding uncertainty relation is always fulfilled.

\end{abstract}

\bigskip


\tableofcontents

\bigskip
\bigskip

\section{Introduction}
\label{sec:intro}


The Shannon information entropy~\cite{Shannon1948comm} of a given probability density $\rho(z)$ is defined as the functional
\begin{equation}
\label{eq:infoS}
S_\rho = - \int \rho (z) \log \rho (z) \, \mathrm d z \,.
\end{equation}
As it is well-known, this entropy essentially measures the total spreading of the probability density and becomes the cornerstone of a huge number of applications of information theory (see, for instance,~\cite{CoverThomas,Gray1990book} and references therein). In particular, this information-theoretic viewpoint can be used in order to analyse the spreading properties of the probability density $\rho (x) = |\psi (x)|^2 $ that characterizes the
stationary states $\psi (x)$ of a given quantum system (see also~\cite{CGAB2021shannon} for a Shannon entropy approach to dynamical stability in classical systems). Moreover, for generic quantum states, $S_\rho$ provides a more appropriate measure of the uncertainty in the position for such a quantum state than the usual Heisenberg uncertainty relation (see~\cite{Hilgevoord2002,ToranzoDehesa2019}). In the same manner, the Shannon entropy $S_\gamma$ for the associated probability density on momentum space $\gamma (p) = |\tilde\psi (p)|^2 $ can be also computed in terms of the momentum representation of the state and, as expected, provides the information concerning the spreading of the momentum distribution for such an state. Moreover, from these two entropies a stronger version of the Heisenberg uncertainty relation was introduced by Bialynicki-Birula and Mycielski~\cite{BM1975} (see also~\cite{Beckner}). For a $N$-dimensional quantum state, this uncertainty relation reads
\begin{equation}
S_\rho + S_\gamma \geq N\,(1 + \log \pi) \, ,
\label{uncerSS}
\end{equation}
which indeed implies the impossibility of getting completely precise information of the quantum state in both position and momentum spaces.

Obviously, the problem of determining in exact form the Shannon entropy of the states for a given quantum system can be only faced provided that system is exactly solvable, {\em i.e.}~when the physical solutions of the corresponding Schr\"odinger equation can be analytically found. The list of quantum systems that belong to this class is a very short one, and their most relevant representatives in the generic $N$-dimensional case are the harmonic oscillator and the Coulomb systems, but we recall that another remarkable class of exactly solvable systems has been recently presented in~\cite{Olendski} and is given by $N$-dimensional Dirichlet and Neumann hyperspherical dots. In particular, the analytical expressions for the Shannon information entropy of the quantum states of the $N$-dimensional harmonic oscillator have been studied in~\cite{YVAS94,VAYD95,DAY1997entropy,MO96,JSR97,DehesaJMP98,ToranzoDehesa2019,ToranzoDehesaRydberg} (see also references therein) where many challenging mathematical problems have been solved.

Taking into account the previous considerations, the aim of this paper is to compute the Shannon entropy for an $N$-dimensional quantum nonlinear oscillator, whose classical version was firstly  introduced in~\cite{BEHR2008PhysicaD}. This system can be interpreted as an exactly solvable deformation (governed by a real and positive parameter $\lambda$) of the $N$-dimensional harmonic oscillator potential that is defined on a very specific space with nonconstant negative curvature, which is just the
$N$-dimensional spherically symmetric generalization~\cite{Ballesteros:2007dq,BEHR2008PhysicaD} of the so-called Darboux surface of type III~\cite{Ko72,KKMW03}. As a consequence, this nonlinear oscillator is known in the literature as the Darboux III oscillator, and its eigenfunctions and eigenvalues can be analytically obtained as smooth $\lambda$-deformations of the ones associated to the $N$-dimensional harmonic oscillator~\cite{BEHR2008PhysicaD,Ballesteros:2007dq, Ballesteros20091219,BEHRR2011quantum,BEHRR2011}.

Therefore, the Darboux III system provides a very distinguished example of exactly solvable $N$-dimensional quantum nonlinear oscillator whose wave functions are amenable to be studied from the information-theoretic viewpoint. Moreover, the deformation approach here presented provides a privileged benchmark in order to analyze the interplay between information entropy and curvature for exactly solvable quantum systems defined on curved spaces, a subject that -to the best of our knowledge- cannot be found in the literature so far (we recall that in~\cite{NPH2020curvatureCK}, information entropies for some quantum states for the free motion on the 2D spherical and hyperbolic spaces with constant curvature have been computed numerically). Nevertheless, we will find that due to the complexity induced by the curvature of the manifold where the system is defined, while the information entropy for the wave functions in the position representation can be analytically computed, this will not be the case in the momentum representation, where numerical methods will be needed in order to compute $S_\gamma$ and to check the uncertainty relation~\eqref{uncerSS}. Moreover, we stress that the current interest in multidimensional harmonic oscillators is outstanding in many different classical and quantum dynamical systems (see~\cite{ToranzoDehesaRydberg} and references therein) and therefore the Darboux III oscillator provides a generalized oscillator model in which  $\lambda$ could be thought of as an effective parameter that can be used to model analytically smooth nonlinear perturbations of all these harmonic phenomena.

The structure of the paper is the following. In the next section the essential features of the Darboux III oscillator, in both its classical and quantum versions, as well as the geometry of its underlying curved space will be shortly reviewed (see~\cite{Ballesteros:2007dq,BEHR2008PhysicaD,Ko72,KKMW03,Ballesteros20091219,BEHRR2011,BEHRR2011quantum} for a detailed account of all the results here sketched). In particular, it will be emphasized that  the quantization of the classical system is by no means unique, since due to the nonvanishing curvature of the underlying manifold the kinetic energy term of the Hamiltonian contains both momenta and position variables, and a precise ordering between them has to be prescribed. In section 3 the information entropy for the one-dimensional Darboux III quantum oscillator will be computed. In position space this will be obtained analytically, and it will be shown that for a given quantum state the information entropy $S_\rho$ increases with $\lambda$, {\em i.e.} with the absolute value of the (negative) curvature. However, in the momentum space representation the Fourier transform of the wave functions cannot be obtained in closed form due to the $\lambda$-deformation. This implies that the corresponding information entropy has to be computed numerically, and now the information entropy $S_\gamma$ becomes smaller for larger values of the deformation parameter $\lambda$. Moreover, for all the excited states the total entropy $S_\rho + S_\gamma$ will be found to decrease when $\lambda$ increases, while the ground state presents the opposite behaviour,  and indeed the uncertainty relation~\eqref{uncerSS} holds in all the cases. Section 4 is devoted to the generalization to $N$-dimensions of the analytical results for the position space representation by making use of hyperspherical coordinates, and special emphasis will be devoted to the analysis of the three-dimensional case due to its physical significance. In this case the entropy on momentum space $S_\gamma$ will be also computed numerically, and the dependence of all the entropies in terms of the curvature parameter $\lambda$  coincides exactly with the previous findings for the one-dimensional Darboux III oscillator. Finally, a concluding section pointing out some remarks and open questions closes the paper. Also, an Appendix contains the numerical values used for the plots presented in the paper.



\section{The Darboux III oscillator}
\label{sec:2}



\subsection{The classical system and the Darboux III space}

The model that we will consider in this paper is an exactly solvable ND quantum nonlinear oscillator whose classical analogue is defined  by the Hamiltonian
\be
{\cal H}(\bq,\bp)={\cal T}(\bq,\bp)+{\cal U}(\bq)=
\frac{\bp^2}{2(1+\la \bq^2)}+\frac{ \omega^2 \bq^2}{2(1+\la \bq^2)}  .
 \label{ac}
\ee
with real parameters $\la\geq0$ and $\om\geq 0$ and where $(\bq,\bp)\in\RR^{2N}$ are conjugate coordinates and momenta. This system was proven to be maximally superintegrable in~\cite{BEHR2008PhysicaD}, and its $(2N-1)$ functionally independent constants of motion are the ones that encode the radial symmetry of the system, namely,
\be
  C^{(m)}=\!\! \sum_{1\leq i<j\leq m} \!\!\!\! (q_ip_j-q_jp_i)^2 , \quad 
 C_{(m)}=\!\!\! \sum_{N-m<i<j\leq N}\!\!\!\!\!\!  (q_ip_j-q_jp_i)^2 , \quad m=2,\dots,N \, ,\label{af}
 \ee
together with the additional  $\lambda$-dependent  integrals
\be
 I_i=p_i^2-\bigl(2\la  {\cal H}(\bq,\bp)-\om^2\bigr) q_i^2 ,\qquad i=1,\dots,N.
\label{ag}
\ee
that contain the Hamiltonian and from which ${\cal H}$ can be written as
\be
{\cal H}=\frac 12 \sum_{i=1}^N I_i \, .
\ee
Therefore, the system defined by ${\cal H}$ can be interpreted as a genuine (maximally superintegrable) $\la$-deformation of the $N$D Euclidean isotropic oscillator with frequency $\om$, since the limit  $\la\to 0$ of (\ref{ac}) yields 
\be
 {\cal H}_0=\frac 12 \bp^2+\frac 12 \om^2\bq^2  .
\ee

From a geometric viewpoint, the term 
\be
{\cal T}(\bq,\bp)=\frac{\bp^2}{2(1+\la \bq^2)}\, ,
\label{kin}
\ee
can be interpreted as the kinetic energy defining the geodesic motion of a particle with unit mass  on  the $N$D radially symmetric generalization~\cite{Ballesteros:2007dq, Ballesteros20091219} of the Darboux surface of type III (see~\cite{Ko72,KKMW03}) whose metric is given by
\be
 \dd s^2= (1+\la \bq^2)\, \dd \bq^2 ,
 \label{ad}
 \ee
in terms of the coordinates $\bq$ of the configuration space of the Hamiltonian system.
It can be checked that~\eqref{ad} defines a $N$-dimensional conformally flat space whose nonconstant scalar curvature  is given by
 \be
R(\bq)=-\la\,\frac{(N-1)\bigl( 2N+3(N-2)\la \bq^2\bigr)}{(1+\la \bq^2)^3} \, ,
 \label{ad1}
 \ee
that always takes negative values in any dimension $N$. Since $\lim_{|\mathbf q|\to \infty}R$ vanishes,  this space is asymptotically flat, and $R(\bq)$ has a minimum at the origin $$R(\mathbf{0})=-2\la N(N-1),$$ which is proportional to the scalar curvature of the $N-$dimensional hyperbolic space. 
 
From the dynamical viewpoint, the central potential 
\be
{\cal U}(\bq)=\frac{\omega^2}{2} \frac{\bq^2}{1+\la \bq^2} \, ,
\label{pot}
\ee was proven in~\cite{Ballesteros20091219,BEHR2008PhysicaD} to be the natural `intrinsic' oscillator potential on the $N$-dimensional Darboux III space. It is worth stressing that in order to preserve the complete integrability (and the exact solvability) of the deformed model,  the modification of the kinetic energy ${\cal T}(\bq,\bp)$ has to be complemented with a suitable $\lambda$-deformation of the oscillator potential given by~\eqref{pot}. In doing so, all bounded trajectories for the classical hamiltonian system defined by ${\cal H}$ turn out to be closed (like in the usual harmonic oscillator) and the Schr\"odinger equation for the quantum version of the system can be also analytically solved by making use of the well-known solution for the $N$-dimensional quantum harmonic oscillator~\cite{BEHRR2011,BEHRR2011quantum}. 

Finally, it is worth stressing that $\cal H$  can be also  interpreted as a nonlinear oscillator on the flat ND Euclidean space but endowed with a position-dependent mass, in which the radially symmetric mass function
$$
 m(\bq)=1+\la \,  \bq^2 \, ,
\label{aadd}
$$
is just the conformal factor of the metric and grows quadratically in terms of $|\bq|$ (see~\cite{Koc, Schd} for certain semiconductor heterostructures described through quadratic mass functions). We emphasize that these two interpretations of the system (motion on a space with variable curvature versus motion of a particle with position-dependent mass on the flat Euclidean space) are fully equivalent, although in this work we will focus on the geometric one.


\subsection{Hyperspherical coordinates and the effective potential}

The radial symmetry of the system suggests the use of hyperspherical coordinates $(r=|\mathbf q|, \bm{\theta} = \{\te_j\}_{j=1,\ldots,N-1})$ given by (see~\cite{BEHRR2011})
\begin{equation}
\begin{split}
q_j &= r \cos \theta_j \prod_{k=1}^{j-1} \sin \theta_k, \qquad \forall j \in \{ 1, \ldots, N-1 \}\\
q_N &= r \prod_{k=1}^{N-1} \sin \theta_k \, .
\end{split}
\end{equation}
These hyperspherical coordinates are defined on the subset of $\mathbb R^N$ given by
\begin{equation}
(r, \theta_1, \ldots, \theta_{N-2},\theta_{N-1} ) \in (0,+\infty) \times [0,\pi] \times \cdots [0,\pi] \times [0,2 \pi] .
\end{equation}

Their corresponding canonically conjugated momenta $(p_r,p_{\te_j})$ are related with the canonical Euclidean momenta by $\mathbf p^2 = p_r^2 + \frac{\mathbf L^2}{r^2}$, where the square of the total angular momenta reads
\be
\mathbf L^2=\sum_{j=1}^{N-1}p_{\te_j}^2\prod_{k=1}^{j-1}\frac{1}{\sin^{2}\te_k}\, .
\ee
In these coordinates the metric (\ref{ad}) now reads
\begin{equation}
\dd s^2= (1+\la r^2)(\dd r^2+r^2\dd\Omega_N^2)
\label{bb}
\end{equation}
where   $\dd\Omega_N^2$  is the metric on the unit $(N-1)$D sphere induced by the usual Euclidean metric on $\mathbb R^N$.

Therefore, the Hamiltonian~\eqref{ac} in hyperspherical coordinates is written as
\begin{equation}
\mathcal H = \frac{ p_r^2}{2(1+\lambda  r^2)} + 
\frac{\mathbf L^2}{2r^2 (1+\lambda r^2)} + \frac{\omega^2  r^2}{2(1+\lambda  r^2)} \, ,
\end{equation}
and the nonlinear oscillator potential
\begin{equation}
\label{eq:U}
U(r) := \frac{\omega^2  r^2}{2(1+\lambda  r^2)} ,
\end{equation}
together with the centrifugal term, defines the effective radial potential for this system
\begin{equation}
\label{eq:Ueff}
U_{\text{eff}}(r) :=\frac{\mathbf L^2}{2r^2 (1+\lambda r^2)} + \frac{\omega^2  r^2}{2(1+\lambda  r^2)} \,.
\end{equation}
Both $U(r) $ and $U_{\text{eff}}(r)$ are plotted in Fig. 1, where it is worth stressing that
\begin{equation}
\lim_{r \to + \infty} U(r) = \lim_{r \to + \infty} U_{\text{eff}}(r) = \frac{\omega^2}{2 \lambda}  \, ,
\end{equation}
which shows that  when $\lambda\neq 0$ the effective potential $U_{\text{eff}}(r) $ saturates for large $r$, thus indicating the `hydrogen-like' nature of the effective potential of the $\lambda$-deformed oscillator~\cite{BEHRR2011}.

\begin{figure}[H]
\begin{center}
\label{fig:spectrum}
\includegraphics[scale=0.6]{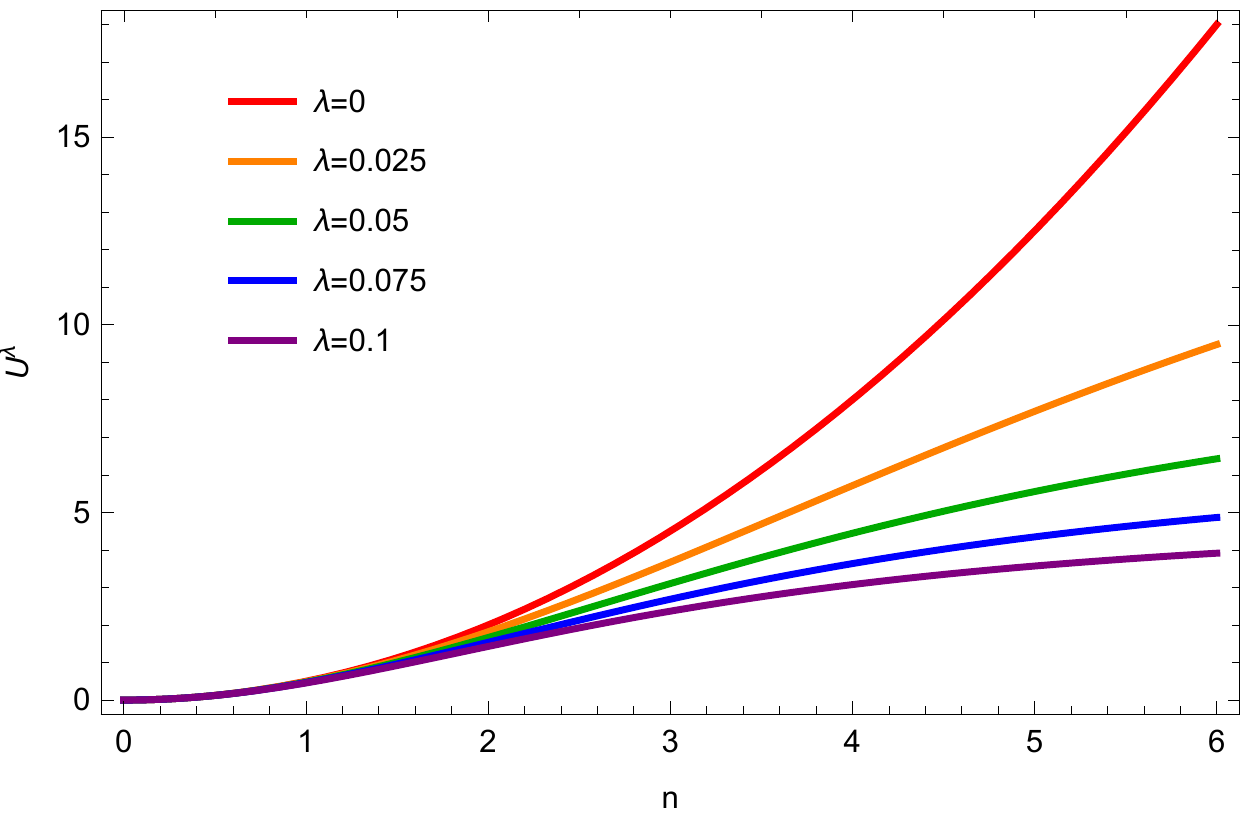}\hspace{1cm}
\includegraphics[scale=0.6]{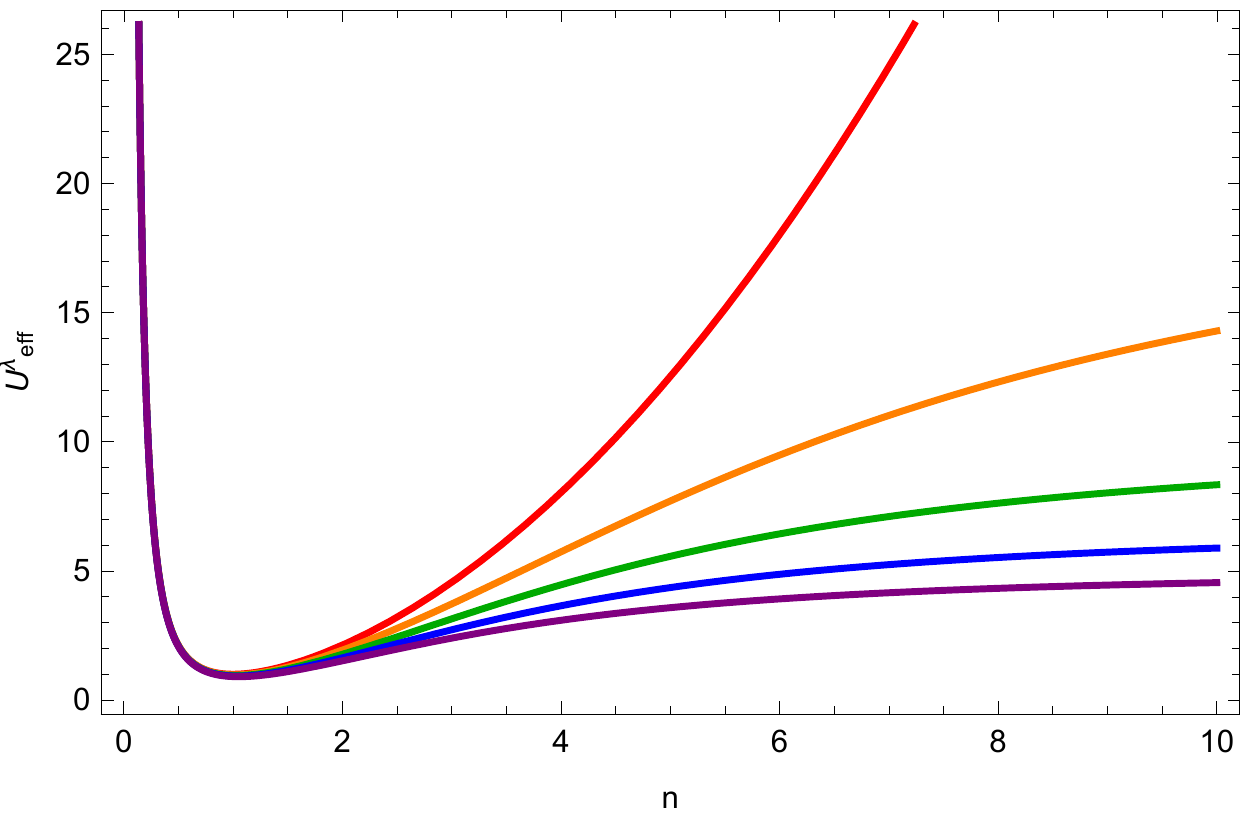}
\caption{Left: Nonlinear oscillator potential $U$ \eqref{eq:U} and effective nonlinear oscillator potential $U_{\text{eff}}$ \eqref{eq:Ueff} for $\omega=1$, $\mathbf L=1$ and different values of $\lambda$.}
\end{center}
\end{figure}


\subsection{The quantum Darboux III oscillator}

As it is well known, the quantization problem for Hamiltonians like~\eqref{ac} whose kinetic energy term~\eqref{kin} contains position-dependent functions, admit different solutions which depend on the ordering prescription chosen for the (non-commuting) position and momenta operators. In general, different orderings lead to quantum Hamiltonians that are related under similarity transformations, therefore their eigenfunctions are related by gauge transformations and their eigenvalues coincide. For a detailed discussion on this issue, we refer to~\cite{BEHRR2011quantum} and references therein.

In the case of the Darboux III Hamiltonian~\eqref{ac}  this problem was analysed in detail in~\cite{BEHRR2011quantum}, from where we sketch in the following the essential results needed for the rest of the paper. Initially, this system was quantized in~\cite{BEHRR2011} by making use of the so-called `Schr\"odinger' quantization prescription, in which the quantum Hamiltonian $\hH$ is given by
\be
 {\hH}= 
 \frac{1}{2(1+\la \hbq^2)}\, \hbp^2+ \  \frac{ \om^2 \hbq^2}{2(1+\la \hbq^2)}  \, ,
 \label{ca}
 \ee
 where the function depending on the coordinates is located at the left within the kinetic energy term.
Then it is straightforward to prove that  $\hH$ commutes with the following observables,
\be
  \hC^{(m)}=\!\! \sum_{1\leq i<j\leq m} \!\!\!\! ( \hq_i \hp_j- \hq_j \hp_i)^2  , \quad 
  \hC_{(m)}=\!\!\! \sum_{N-m<i<j\leq N}\!\!\!\!\!\!  ( \hq_i \hp_j- \hq_j \hp_i)^2  , \quad m=2,\dots,N;
 \ee
\vskip-0.25cm
 \be
  \hI_i= \hp_i^2- 2\la  \hq_i^2 { \hH}( \hbq, \hbp)+ \om^2 \hq_i^2 ,\qquad i=1,\dots,N;
\ee
where $\hC^{(N)}=\hC_{(N)}$ and ${\hH}=\frac 12 \sum_{i=1}^N \hI_i$.  In this setting, each of the three  sets $\{{\hH},\hC^{(m)}\}$,  
$\{{\hH},\hC_{(m)}\}$ ($m=2,\dots,N$) and   $\{\hI_i\}$ ($i=1,\dots,N$) is  formed by $N$ commuting observables. Moreover, it can be proven that $\hH$ is (formally) self-adjoint on the Hilbert space $L^2(\RR^N,(1+\la\bq^2){\dd}\bq  )$, endowed with the scalar product
\be
\langle \Psi | \Phi \rangle_\la := \int_{\RR^N} \overline{{\Psi}(\bq)} \Phi(\bq)(1+\la\bq^2){\dd}\bq \, ,
\ee
in which the conformal factor of the metric (or mass function) plays an outstanding role.

Generic eigenfunctions were then found in~\cite{BEHRR2011},  and turn out to be formally analogous to the ones for the $N$-dimensional isotropic oscillator but provided that the frequency of the oscillator transforms into an energy-dependent function. The explicit expression for the eigenvalues of~\eqref{ac} can be also sraightforwardly obtained and reads
\begin{equation}
E_n^\lambda= -\hbar^2 \lambda\left(n + \frac{N}{2}\right)^2 + \hbar \left(n+\frac N 2\right ) \sqrt{\hbar^2 \lambda^2 \left(n+\frac{N}{2}\right)^2+ \om^2 }  \, ,
\label{re}
\end{equation}
where we see that the flat limit $\lambda\to 0$ gives the spectrum of the flat isotropic oscillator, 
the limit $n\to \infty$ leads to the upper energy limit $\om^2/2\lambda$ for bounded states, and the degeneracy of the model is exactly the same as in the $N$D isotropic oscillator, a feature that is again a signature of its maximal superintegrability. Finally, the continuous spectrum of $\hH$ is given by $[\frac{\om^2}{2\la },\infty)$. 

However, as we will see in the following section, in order to be able to compute the information entropy for this model both in the position and in the momentum spaces, a quantization prescription for which the quantum Darboux III Hamiltonian  is formally self-adjoint on the space $L^2(\mathbb R^N)$, with the usual inner product 
\begin{equation}
\langle f,g \rangle = \int_{- \infty}^\infty \overline{f(\mathbf q)} g(\mathbf q) \mathrm d \mathbf q \, ,
\label{innerflat}
\end{equation}
would be of the outmost relevance. Remarkably enough, such a quantization was also given in~\cite{BEHRR2011quantum}, and is given by a quantum Hamiltonian of the form~\cite{mass2}
\be
  \hat{\cal H}_{\rm PDM}(\hat\bq,\hat\bp)=\frac { 1} 2\, \hbp\, \frac{1}{(1+\la \hbq^2)}\, \hbp
  + \  \frac{ \om^2 \hbq^2}{2(1+\la \hbq^2)} =-\frac {\hbar^2} 2 \nabla   \frac{1}{(1+\la \bq^2)} \nabla  + \frac{\omega^2 \mathbf{q}^2}{2(1+\lambda \mathbf{q}^2)} \, ,
\label{oa0}
\ee
which is is based on the  symmetrization of the kinetic energy term that is often used in the Condensed Matter literature. After reordering terms in order to make connection with the Hamiltonian for the quantization prescription~\eqref{ca} and afterwards by adding suitable potential terms depending on $\hbar$, we are lead to the so-called Transfomed-Position-Dependent-Mass (TPDM) Hamiltonian~\cite{BEHRR2011quantum}
\begin{equation}
\label{eq:HND}
\hat{ \mathcal H}_{\mathrm{TPDM}}  = \frac{1}{2(1+\lambda \, \hat{\mathbf{q}}^2)}\hat{\mathbf p}^2 +\frac{\omega^2 \hat{\mathbf{q}}^2}{2(1+\lambda \, \hat{\mathbf{q}}^2)} + \frac{i \hbar \lambda \hat{\mathbf{q}}}{2(1+\lambda \,\hat{\mathbf{q}}^2)^2}\hat{\mathbf p} + \frac{\hbar^2 \lambda (1-2 \, \lambda \, \hat{\mathbf{q}}^2)}{2(1+\lambda \, \hat{\mathbf{q}}^2)^3} \, .
\end{equation}
This Hamiltonian can be shown to be formally self-adjoint with respect to~\eqref{innerflat}, its eigenvalues coincide with~\eqref{re}, and the corresponding eigenfunctions can be found and will be explicitly given in the following section. It is worth stressing that the two last terms in~\eqref{eq:HND} do not appear in~\eqref{ca}, and are essential in order to define a self-adjoint operator with respect to~\eqref{innerflat} that can be obtained from~\eqref{ca} through a similarity transformation. Obviously, these two terms vanish in the classical limit $\hbar\to 0$ and reflect the possible multiplicity of quantizations for a given classical Hamiltonian system that lead to the same spectrum. A detailed account of all these assertions is provided in~\cite{BEHRR2011quantum}.

With all these results at hand, the complete analysis of the Shannon information entropy of this quantum Darboux III oscillator can be fully performed, and its behavior in terms of the deformation parameter $\lambda$ can be studied.


\section{Shannon entropy for the one-dimensional Darboux III oscillator}


\begin{figure}[t]
\begin{center}
\includegraphics[scale=0.6]{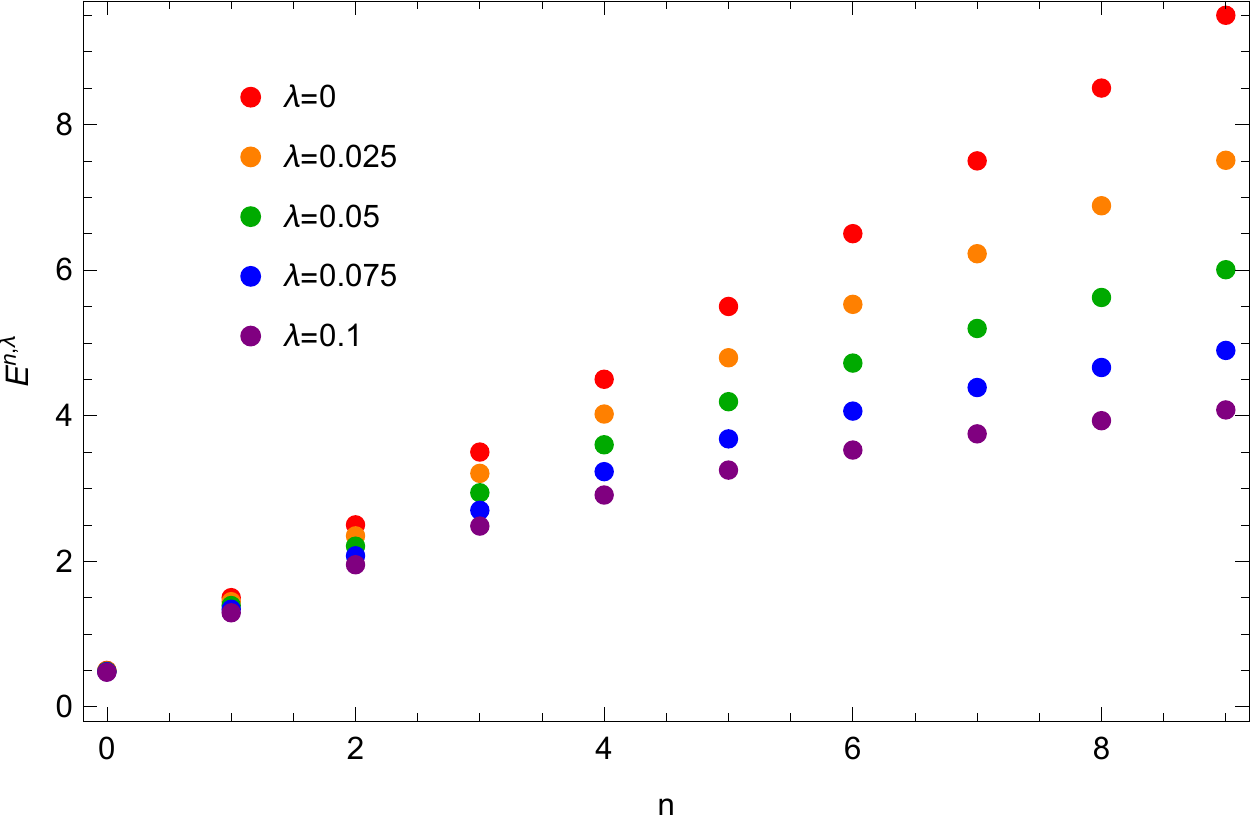}\hspace{1cm}
\includegraphics[scale=0.6]{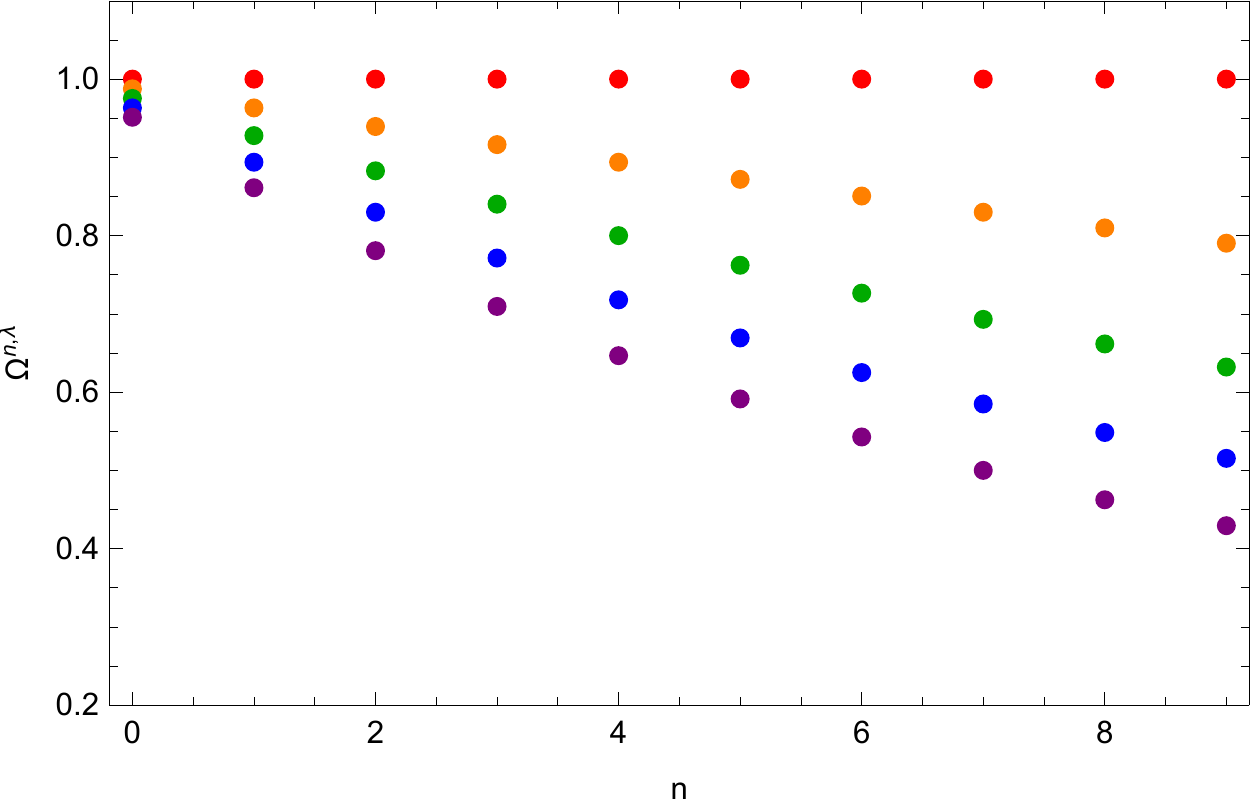}
\caption{\label{fig:spectrum}
Discrete spectrum $E_n^\lambda$ (left) and frequencies $\Omega_n^\lambda$ (right) for the $n=0,\ldots,15$ states and different values of $\lambda$.}
\end{center}
\end{figure}

Let us start by computing the Shannon information entropy \eqref{eq:infoS} of the eigenstates corresponding to the system defined by \eqref{eq:HND} in the one-dimensional case. We thus need the probability density associated to each eigenstate of the quantum system, and therefore to explicitly compute such eigenfunctions. 

\subsection{Probability density}

The one-dimensional version of the Hamiltonian \eqref{eq:HND} discussed in Section \ref{sec:2} reads
\begin{equation}
\label{eq:H1D}
\hat{ \mathcal H}_{\mathrm{TPDM}} = \frac{1}{2(1+\lambda \, \hat x^2)}\hat p^2 +\frac{\omega^2 \hat x^2}{2(1+\lambda \, \hat x^2)} + \frac{i \hbar \lambda \hat x}{2(1+\lambda \, \hat x^2)^2}\hat p + \frac{\hbar^2 \lambda (1-2 \, \lambda \, \hat x^2)}{2(1+\lambda \, \hat x^2)^3}
\end{equation}
which is formally self-adjoint on $L^2(\mathbb R)$ endowed with the usual inner product 
\begin{equation}
\langle f,g \rangle = \int_{- \infty}^\infty \overline{f(x)} g(x) \mathrm d x .
\end{equation}
Therefore all eigenvalues $E_n^\lambda$ are real, and they take the form \cite{BEHRR2011quantum}
\begin{equation}
E_n^\lambda = - \hbar^2 \left( n + \frac{1}{2} \right)^2 + \hbar \left( n + \frac{1}{2} \right) \sqrt{\hbar^2 \lambda^2 \left( n + \frac{1}{2} \right)^2 + \omega^2} \, ,
\end{equation}
where $n=0,1,2,\dots$ is the only quantum number in this one-dimensional system. These energies are plotted in Figure \ref{fig:spectrum}  (see Table \ref{table:spectrum} for the corresponding numerical values) for different values of $\lambda$ and in the limit $\lambda\to 0$ we get the energies for the one-dimensional oscillator states $E_n^0 =\hbar \omega \left( n + \frac{1}{2} \right)$. In \cite{BEHRR2011quantum} the eigenfunctions for~\eqref{eq:H1D} were proven to be 
\begin{equation}
\label{eq:psin}
\psi_n^\lambda (x) = \left(\frac{\Omega_n^\lambda}{\pi}\right)^\frac{1}{4} \sqrt{\frac{1}{2^n n!}} \sqrt{\frac{1}{1+\left( n + \frac{1}{2}\right) \frac{\lambda}{\Omega_n^\lambda}}} \sqrt{1+\lambda x^2} \; e^{-\frac{\Omega_n^\lambda x^2}{2}} H_n \left( \sqrt{\Omega_n^\lambda} \,x \right) \, ,
\end{equation}
where the energy-dependent frequencies $\Omega_n^\lambda$ of this quantum nonlinear oscillator are given by 
\begin{equation}
\Omega_n^\lambda := \sqrt{\omega^2 - 2 \lambda E_n^\lambda} .
\label{nen}
\end{equation}
Therefore, eigenstates~\eqref{eq:psin} are essentially the ones for a quantum oscillator with frequancy $\Omega_n^\lambda$ multiplied by the extra factor $\sqrt{1+\lambda x^2}$ which encodes the role played by the underlying curved space.
In the limit  of vanishing curvature $\lambda \to 0$ we have have that $\Omega_n^\lambda\rightarrow \omega $ and thus we exactly recover the well-known eigenfunctions of the one-dimensional harmonic oscillator
\begin{equation}
\psi_n^0 (x) = \lim_{\lambda \to 0} \psi_n^\lambda (x) = \left(\frac{\omega}{\pi}\right)^\frac{1}{4} \sqrt{\frac{1}{2^n n!}} \; e^{-\frac{\omega x^2}{2}} H_n \left( \sqrt{\omega} x \right) .
\end{equation}

The probability density associated with these states is straightforwardly given by
\begin{equation}
\label{eq:rhon}
\rho_n^\lambda (x) = |\psi_n^\lambda (x)|^2 = \left(\frac{\Omega_n}{\pi}\right)^\frac{1}{2} \frac{1}{2^n n!} \frac{1}{1+\left( n + \frac{1}{2}\right) \frac{\lambda}{\Omega_n}} (1+\lambda x^2) \; e^{-\Omega_n x^2} H_n^2 \left( \sqrt{\Omega_n^\lambda} x \right) ,
\end{equation}
and the limit $\lambda \to 0$ gives
\begin{equation}
\rho_n^0 (x) = \left(\frac{\omega}{\pi}\right)^\frac{1}{2} \frac{1}{2^n n!} \; e^{-\omega x^2} H_n^2 \left( \sqrt{\omega} x \right) .
\end{equation}
Figure \ref{fig:rho_space} shows this probability density $\rho_n^\lambda (x)$ for the ground and some excited states  by considering several values of the curvature parameter $\lambda$. It can be clearly appreciated that as far as $n$ grows the curvature $\lambda$ strongly increases the spreading properties of the wave functions of the system.

\subsection{Shannon entropy in position space}

We now compute the information entropy \eqref{eq:infoS} for the one-dimensional Darboux III oscillator from its probability density $\rho(x)$ given by \eqref{eq:rhon}. Similarly to the case of the harmonic oscillator (see for instance~\cite{DAY1997entropy}), $S_\rho$ can be completely written in terms of certain integrals involving Hermite polynomials~\cite{Hermite2009,OMS2009atlas}.

In our case, in order to be able to obtain a closed expression of  the information entropy for arbitrary values of the quantum number $n$ and the curvature parameter $\lambda$, we will need to compute certain integrals involving the square of the Hermite polynomials multiplied by certain  polynomials. Recall that the Hermite polynomials are orthogonal with respect to the measure $\mathrm d \mu(z) = e^{-z^2} \mathrm d z$ on $\mathbb R$, i.e.
\begin{equation}
\int_{- \infty}^\infty H_n (z) H_m (z) e^{-z^2} \mathrm d z = 0  \, ,
\end{equation}
if $m \neq n$, and here we will make use of a normalization such that
\begin{equation}
\int_{- \infty}^\infty H_n^2 (z) \, e^{-z^2} \mathrm d z = \sqrt{\pi} \, 2^n \, n!  \, ,
\end{equation}
for all $n \in \mathbb N$. Since $z^{2 m+1} $ is an odd function, we have that
\begin{equation}
\int_{- \infty}^\infty z^{2 m+1} H_n^2 (z) \, e^{-z^2} \mathrm d z = 0 .
\end{equation}
Now, by using the previous relations and the well-known recurrence formula 
\begin{equation}
H_{n+1} (z) = 2 \, z \, H_n (z) - 2 \, n \, H_{n-1} (z) ,
\end{equation}
it can be shown that 
\begin{equation}
\int_{- \infty}^\infty z^2 \, H_n^2 (z) \, e^{-z^2} \mathrm d z = \sqrt{\pi} \, 2^n \, n!  \left( n +\frac{1}{2} \right) ,
\end{equation}
together with 
\begin{equation}
\int_{- \infty}^\infty z^4 \, H_n^2 (z) \, e^{-z^2} \mathrm d z = \sqrt{\pi} \, 2^n \, n!  \left( \frac{3 n^2}{2} + \frac{3 n}{2} + \frac{3}{4} 
\right) \, ,
\end{equation}
for all $n \in \mathrm N$.




\begin{figure}[H]
\begin{center}
\includegraphics[scale=0.6]{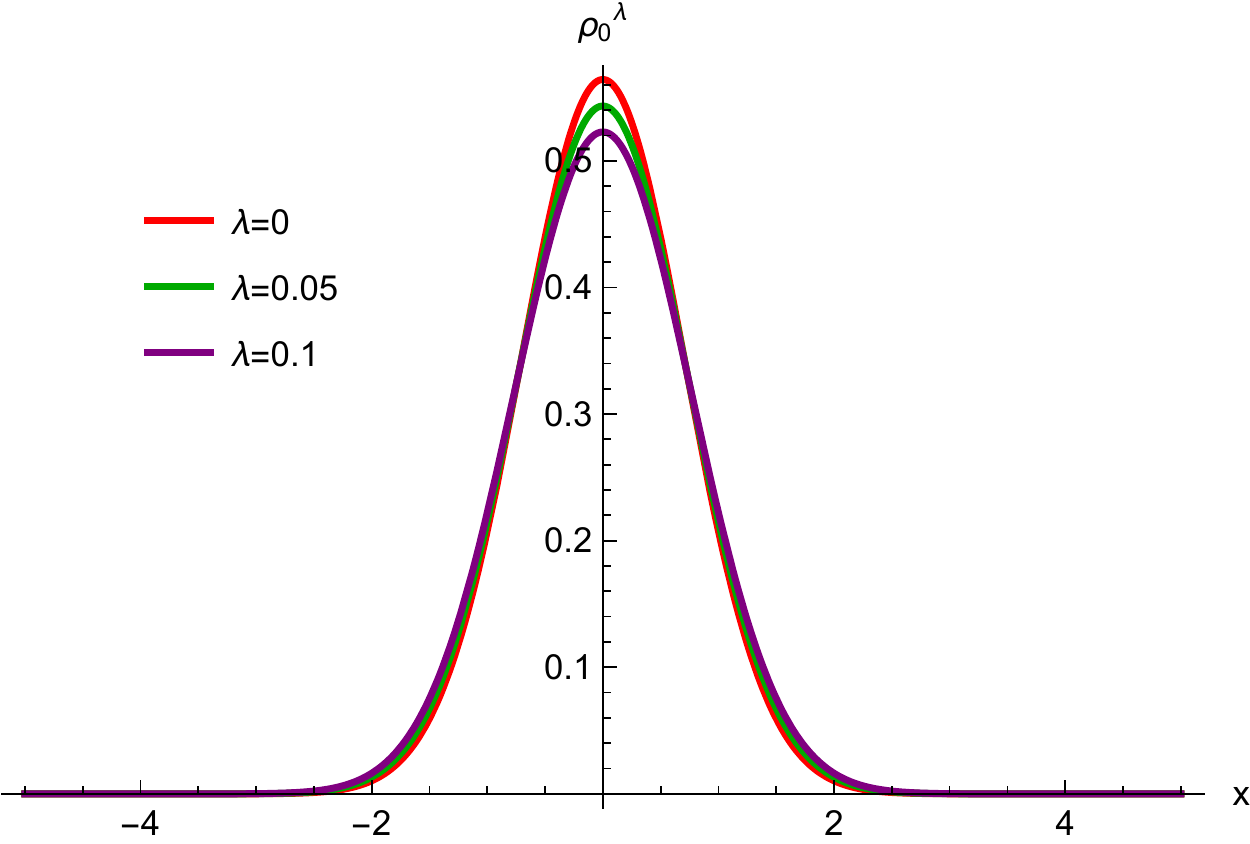}
\includegraphics[scale=0.6]{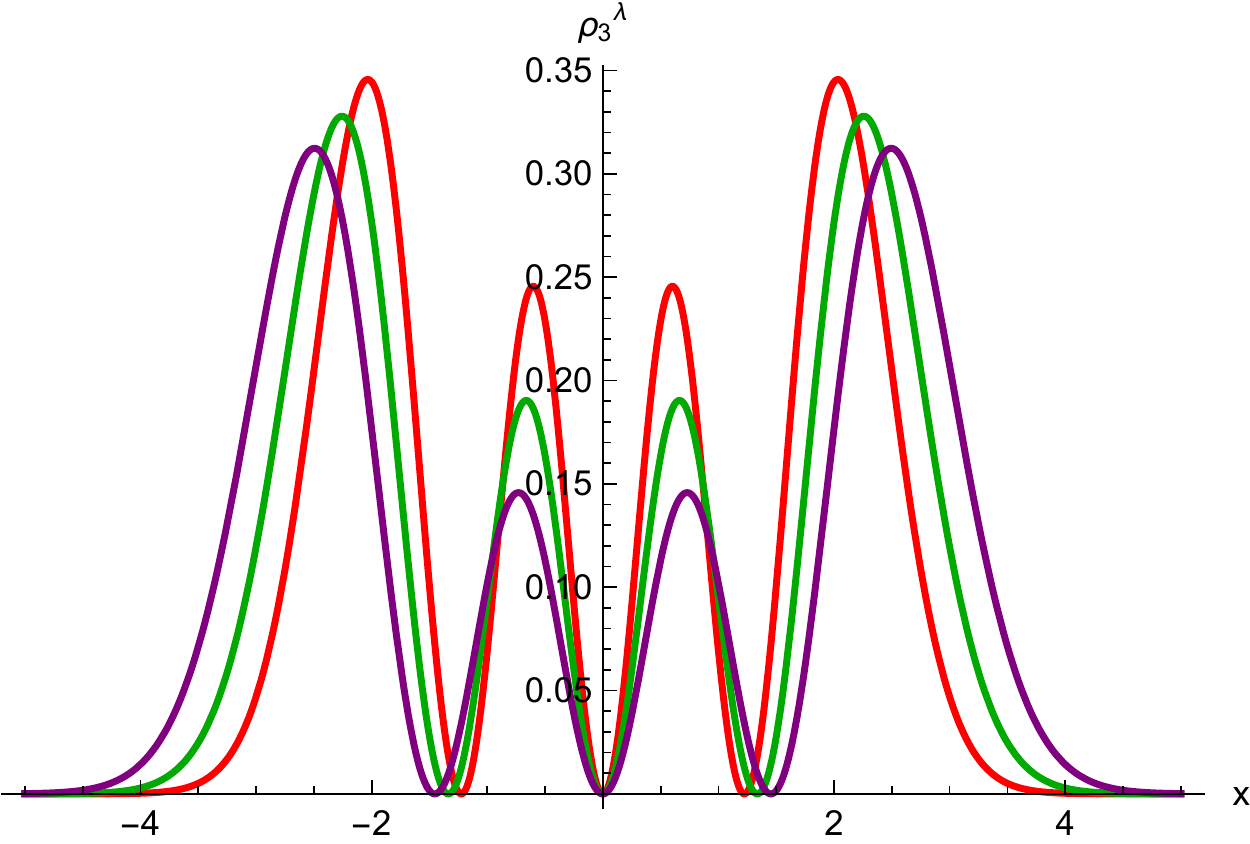} \\
\includegraphics[scale=0.6]{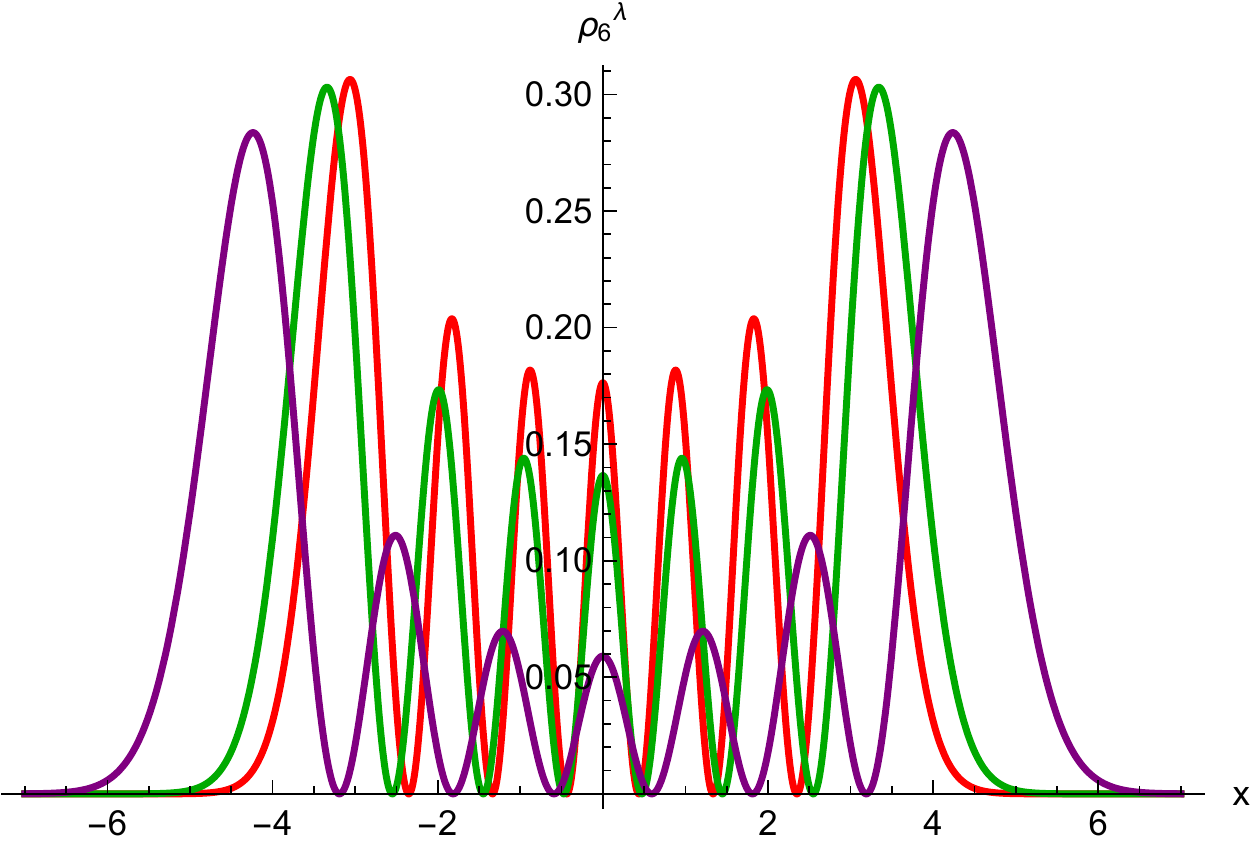} 
\includegraphics[scale=0.6]{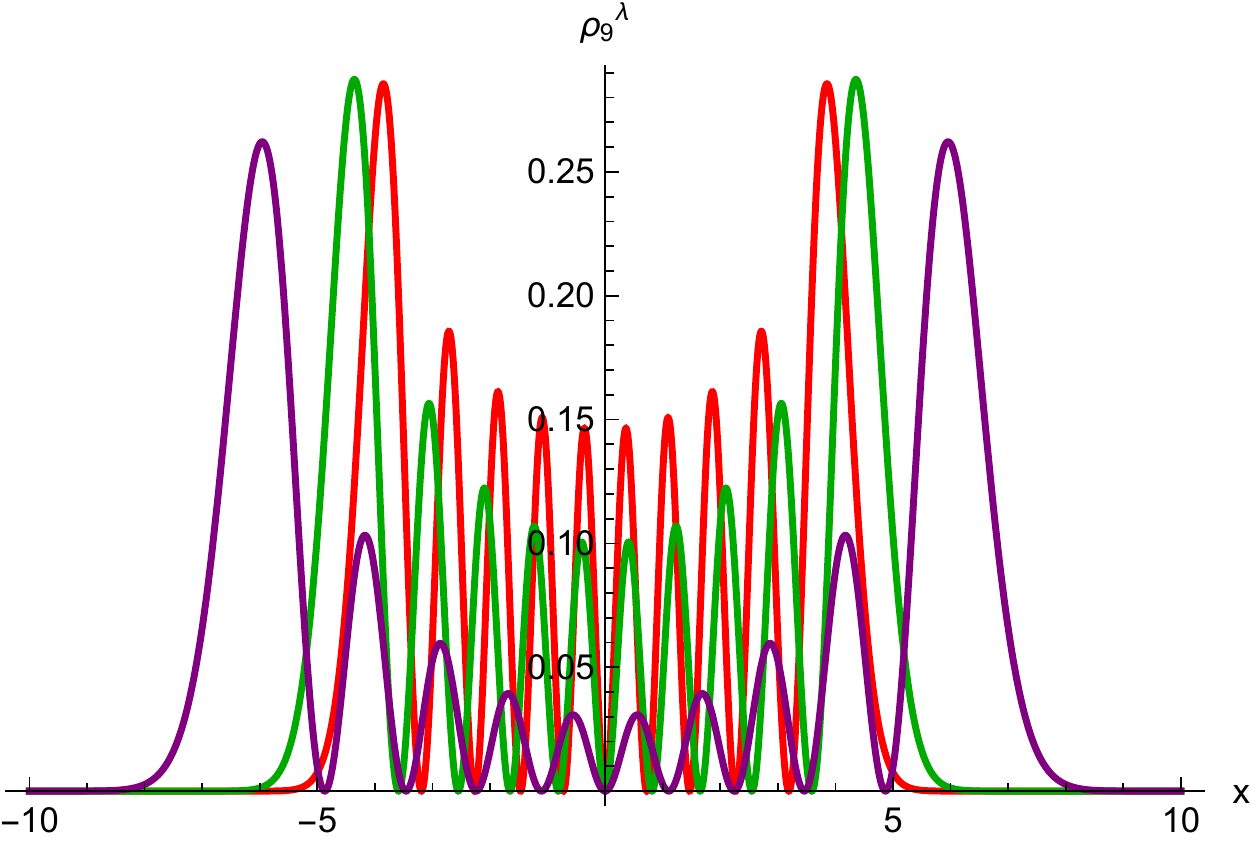}
\caption{Plot of the probability density $\rho_n^\lambda(x)$ for the $n=0,3,6,9$ states and different values of $\lambda$.}
\label{fig:rho_space}
\end{center}
\end{figure}


With all these results at hand, the information entropy $S_\rho^{n,\lambda}$ on the position space  of the eigenstate $\psi_n^\lambda(x)$ of the one-dimensional Darboux III oscillator can be written as
\begin{equation}
\label{eq:entropy1Dspace}
\begin{split}
S_\rho^{n,\lambda} &= - \frac{1}{2} \log \Omega_n^\lambda + \log \left( \sqrt{\pi} \, 2^n \, n! \right) + \log \left( 1+ \left( n + \frac{1}{2} \right) \frac{\lambda}{\Omega_n^\lambda} \right) \\
&+\frac{1}{1+ \left( n + \frac{1}{2} \right) \frac{\lambda}{\Omega_n^\lambda}} \left[ \left( n + \frac{1}{2} + \frac{\lambda}{\Omega_n^\lambda} \left( \frac{3n^2}{2} + \frac{3n}{2} + \frac{3}{4}\right) \right) - \frac{1}{\sqrt{\pi} \, 2^n \, n!} \, \mathcal I^{\frac{\lambda}{\Omega_n^\lambda}} \right] ,
\end{split}
\end{equation}
where  the symbol $\mathcal I^\alpha$ is defined as the following integral
\begin{equation}
\mathcal I^\alpha := \int_{- \infty}^\infty \left( 1 + \alpha \, z^2 \right) e^{-z^2} H_n^2 (z) \log \left( \left( 1 + \alpha \, z^2 \right) H_n^2 (z) \right) \mathrm d z \, .
\end{equation}
It can be straightforwardly checked that in the limit $\lambda\to 0$ of vanishing curvature we recover the results obtained in \cite{DAY1997entropy} for the harmonic oscillator, namely
\begin{equation}
S_\rho^{n,0} = - \frac{1}{2} \log \omega + \log \left( \sqrt{\pi} \, 2^n \, n! \right)  + n + \frac{1}{2} - \frac{1}{\sqrt{\pi} \, 2^n \, n!} \, \mathcal I^0 .
\end{equation}

\begin{figure}[htbp]
\begin{center}
\includegraphics[scale=0.7]{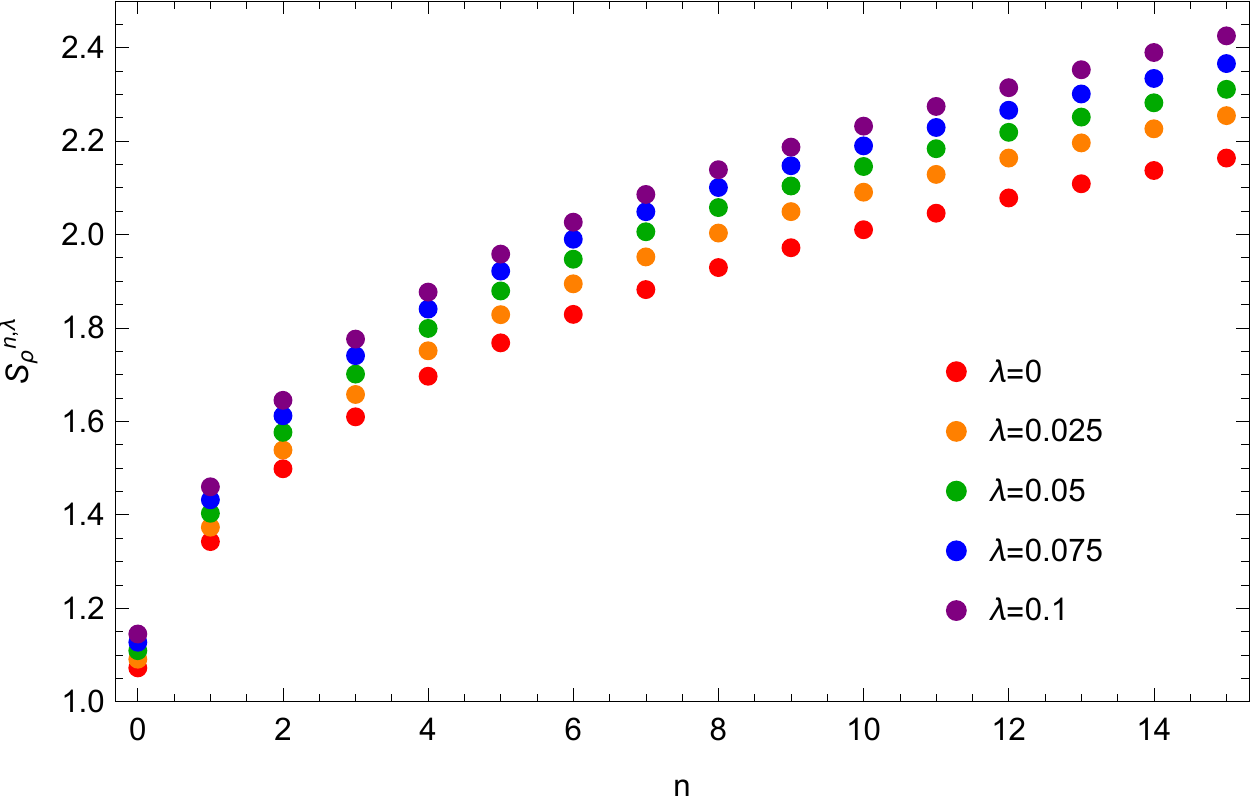}
\caption{Entropies $S_\rho^{n,\lambda}$ for the $n=0,\ldots, 15$ states and different values of $\lambda$.}
\label{fig:entropiesspace}
\end{center}
\end{figure}

The values of the entropy \eqref{eq:entropy1Dspace} in position space for the states with $n=0,\ldots,15$ are 
contained in Table \ref{table:entropiesspace} and have been plotted in Figure \ref{fig:entropiesspace}. It can be clearly appreciated that the spreading induced by the curvature implies that the information entropy of a given eigenstate grows with the parameter $\lambda$. It is interesting to compare this behavior to the one shown in Figure \ref{fig:rho_space}, where the dependence of the probability densities with $\lambda$ is expressed. In particular, Figure \ref{fig:rho_space} shows how differences in $\rho_n^\lambda(x)$ for different values of the curvature grow for states with higher $n$, and this behavior is translated to Figure \ref{fig:entropiesspace} where the differences in entropies for different curvatures also grow as $n$ grows. Summarizing, the curvature parameter $\lambda$ has a ``delocalizing'' effect that is translated in terms of the information entropies, and this effect is additional to the usual one in the harmonic oscillator where the Shannon entropy grows with $n$.

\subsection{Shannon entropy in momentum space}

\begin{figure}[H]
\begin{center}
\includegraphics[scale=0.6]{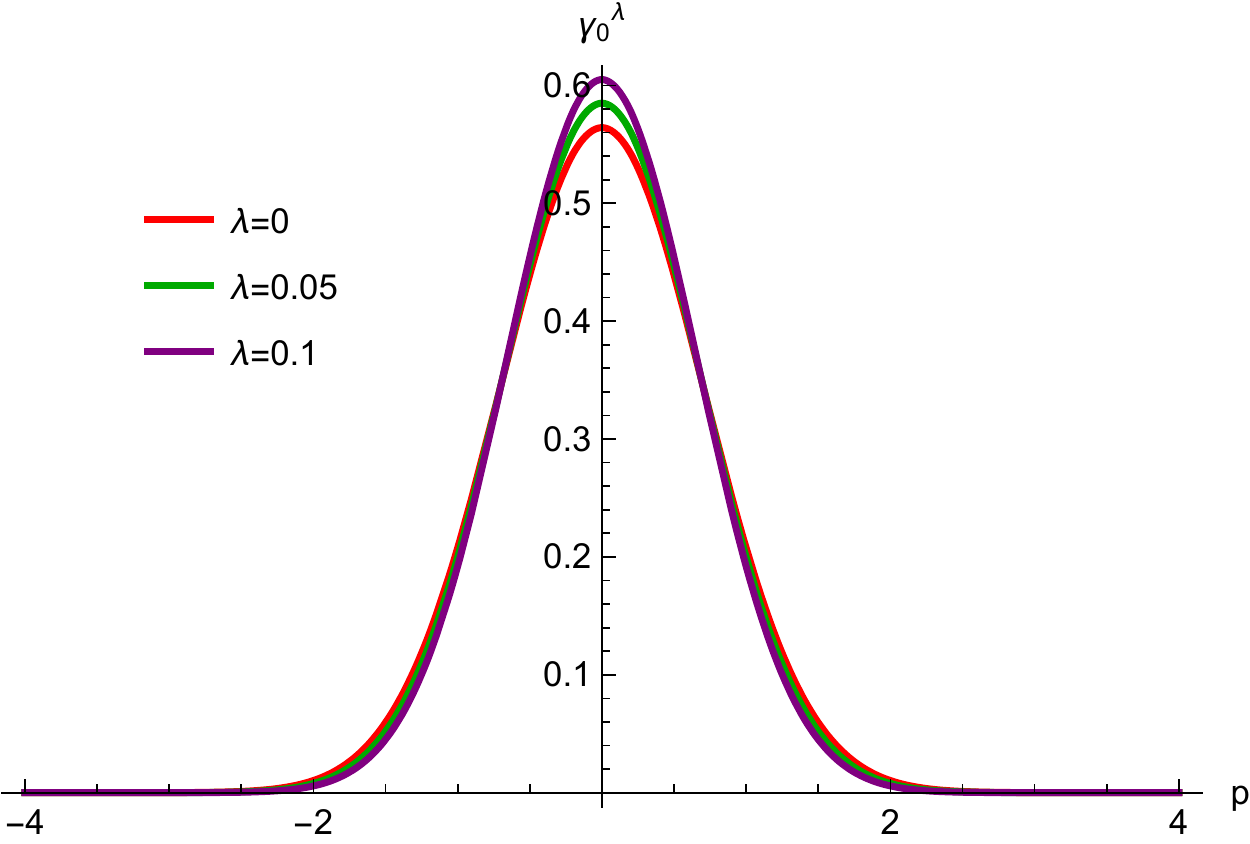}
\includegraphics[scale=0.6]{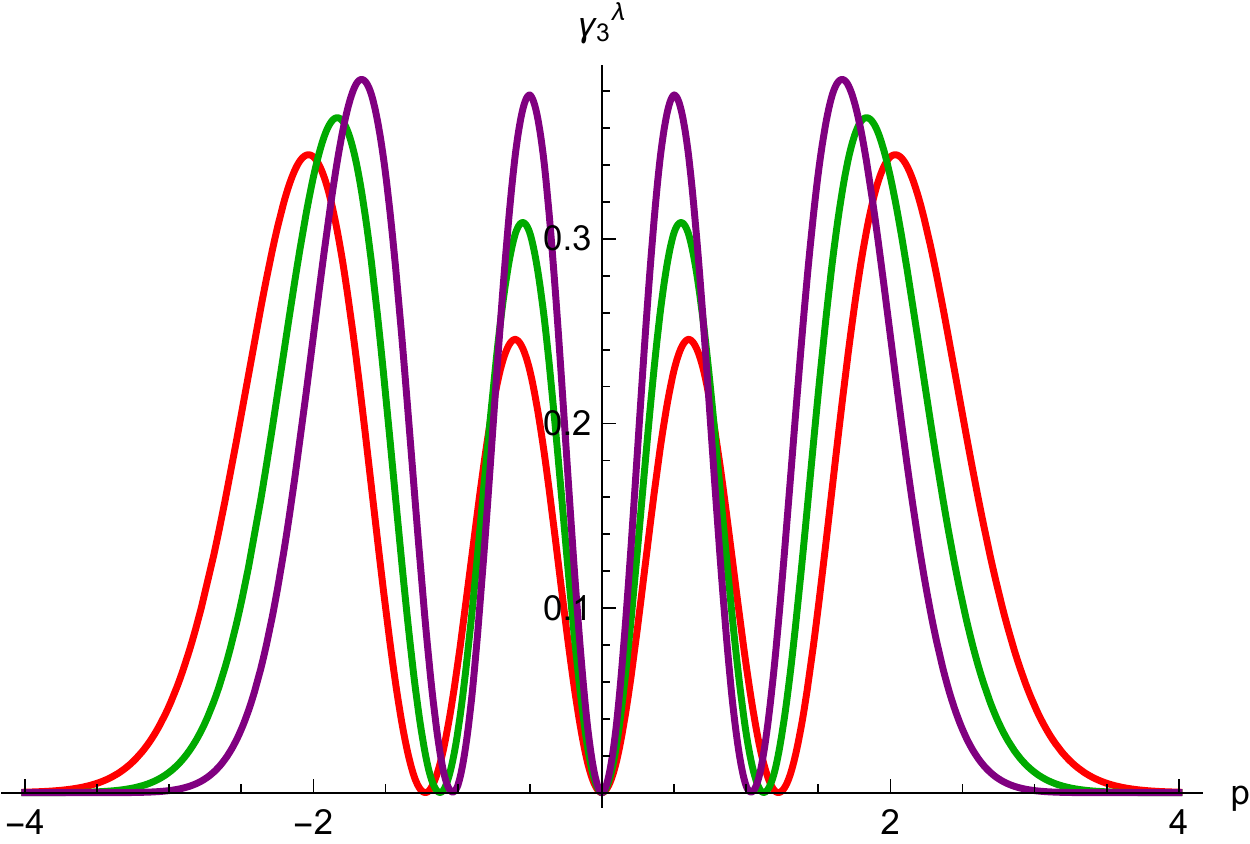} \\
\includegraphics[scale=0.6]{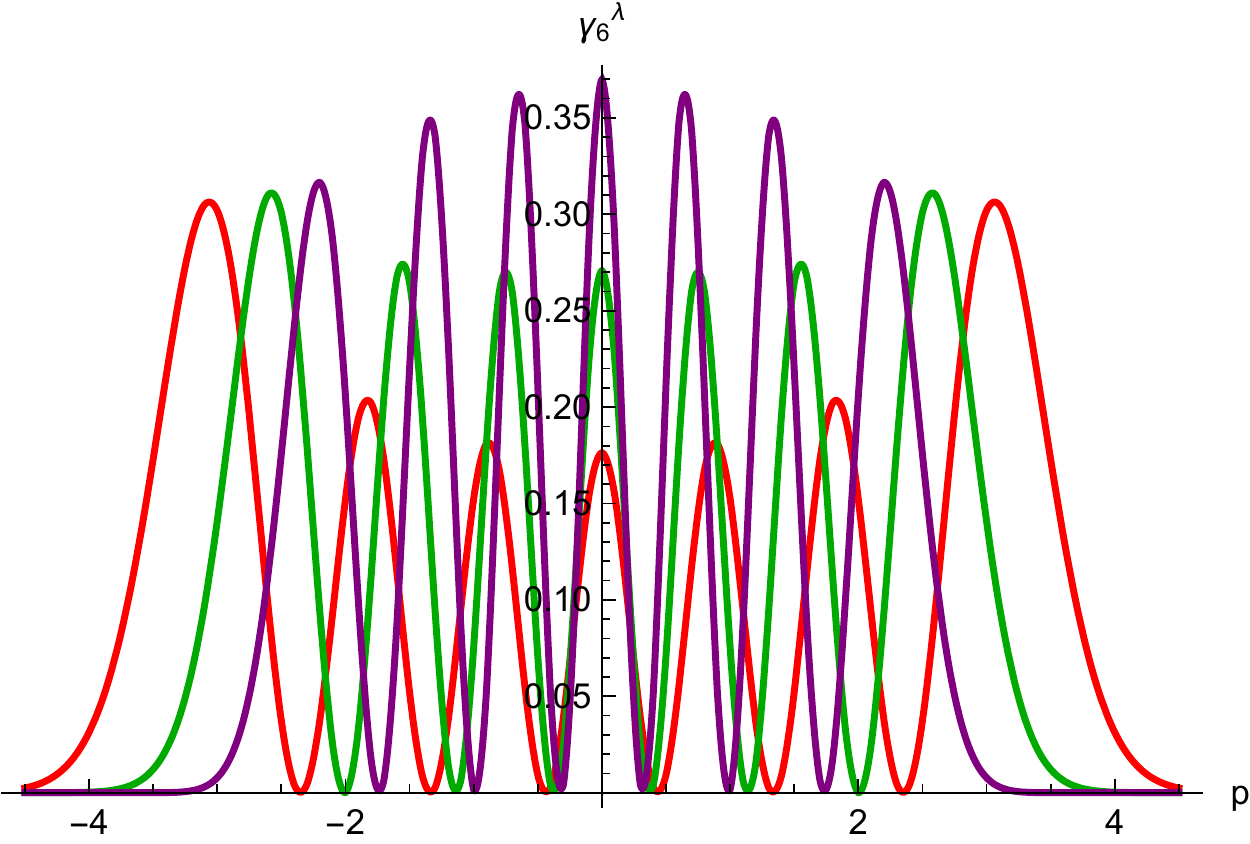} 
\includegraphics[scale=0.6]{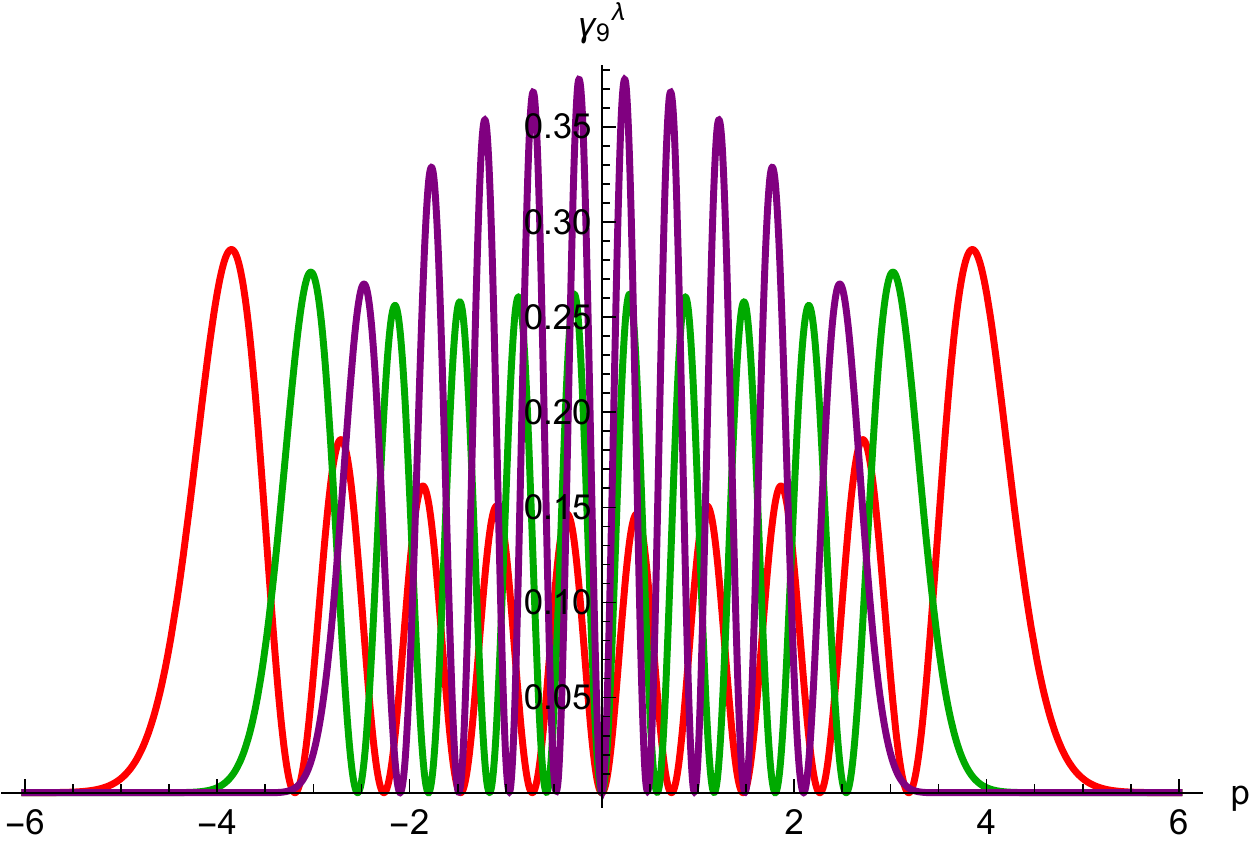}
\caption{\label{fig:gamma1D} Plots of the probability density $\gamma_n^\lambda(p)$ for the $n=0,3,6,9$ states and different values of $\lambda$.}
\end{center}
\end{figure}

In order to compute the information entropy on momentum space we need to compute the Fourier transform \red{of}  the wave-functions \eqref{eq:psin}. In contradistinction to the usual harmonic oscillator in flat space, the Fourier transform for the eigenstates of the Darboux III oscillator
\begin{equation}
\label{eq:psip1D}
\widetilde{\psi_n^\lambda} (p) = \mathcal F \left\{ \psi_n^\lambda (x) \right\} (p) = \frac{1}{\sqrt{2 \pi}} \int_{-\infty}^\infty e^{i p x} \,  \psi_n^\lambda (x) \, \mathrm d x  \, ,
\end{equation}
cannot be expressed as an analytical function. Therefore our results will rely on the numerical computation of \eqref{eq:psip1D}. Indeed, we have that 
\begin{equation}
\label{eq:gamma1D}
\gamma_n^\lambda (p) = |\widetilde{\psi_n^\lambda} (p)|^2  \, ,
\end{equation}
and in the limit $\lambda \to 0$ we do recover the well-known probability density in momentum space
\begin{equation}
\gamma_n^0 (p) = \lim_{\lambda \to 0} \gamma_n^\lambda (p) = | \widetilde{\psi_n^0} (p) |^2 = \left(\frac{1}{\omega\pi}\right)^\frac{1}{2} \frac{1}{2^n n!} \; e^{-\frac{p^2}{\omega}} H_n^2 \left( {\frac{p}{\sqrt\omega}} \right) .
\end{equation}
In Figure \ref{fig:gamma1D} it is shown how the probability density $\gamma_n^\lambda (p)$ in momentum space \eqref{eq:gamma1D} varies with the curvature $\lambda$. It is interesting to note that the qualitative result is the opposite to the one in position space, namely, higher values of $\lambda$ forces the eigenstate \eqref{eq:psip1D} to be more localized in momentum space. 

\begin{figure}[H]
\begin{center}
\includegraphics[scale=0.7]{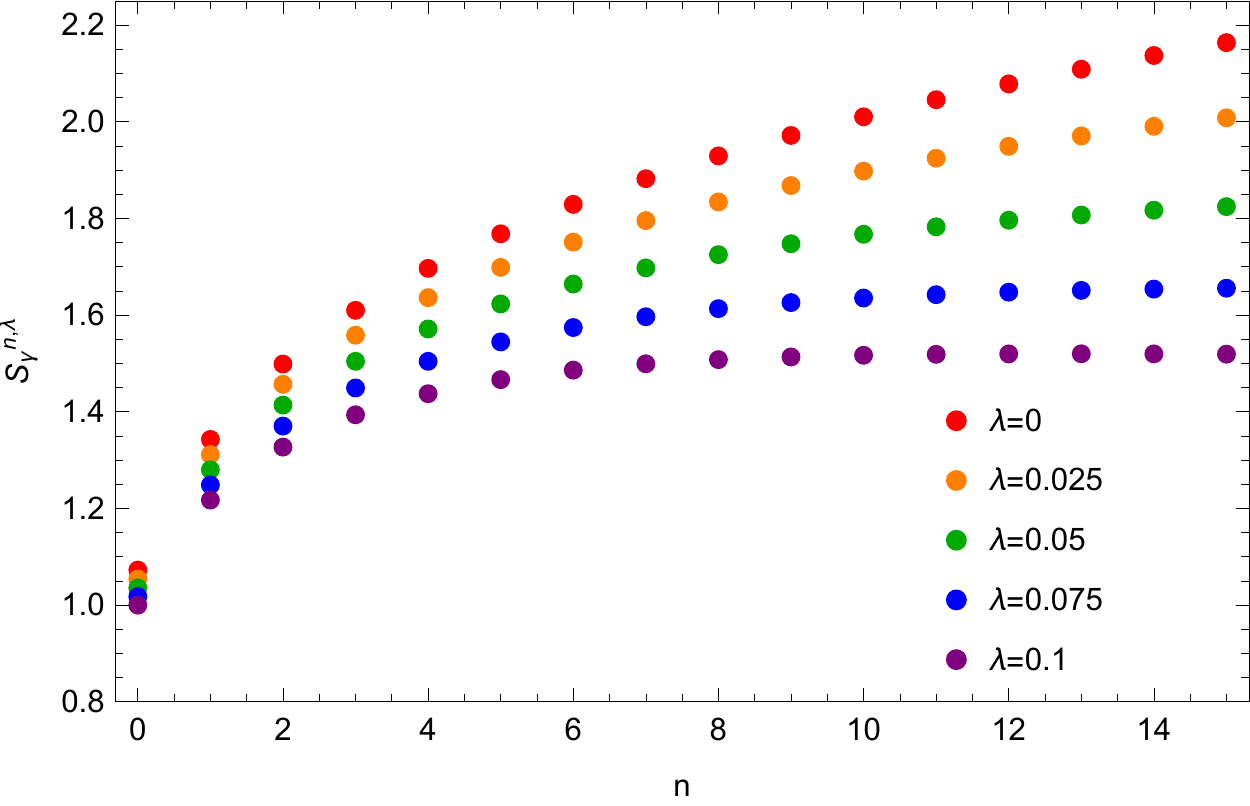}
\caption{\label{fig:entropiesmomentum} Entropies $S_\gamma^{n,\lambda}$ for the states with $n=0,\ldots, 15$ and different values of $\lambda$.}
\end{center}
\end{figure}

This behavior is clearly appreciated in Figure \ref{fig:entropiesmomentum} (data are provided in Table \ref{table:entropiesmomentum}), where greater values of $\lambda$ are associated with smaller values of $S_\gamma$ for each fixed quantum state $\psi_n^\lambda (p)$.

\begin{figure}[H]
\begin{center}
\includegraphics[scale=0.7]{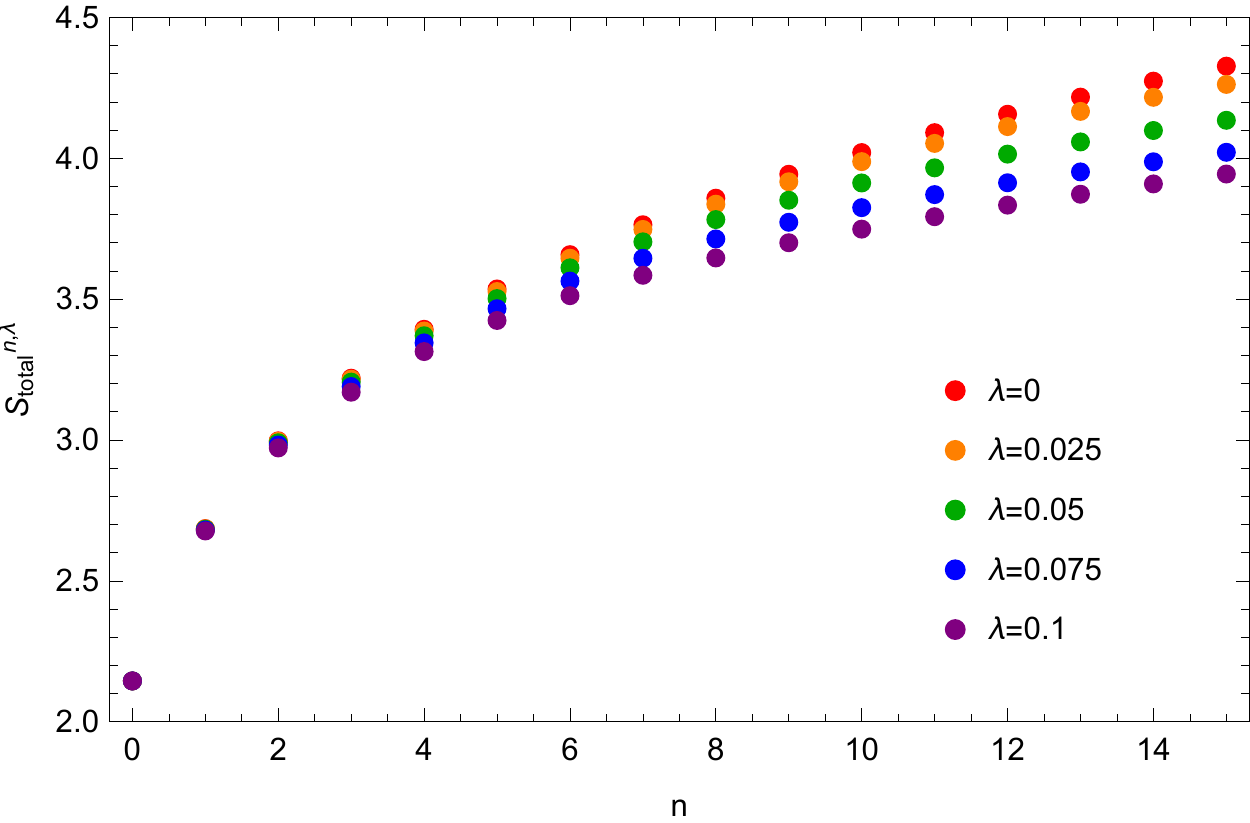}
\caption{ \label{fig:entropiessum} $S_\rho^{n,\lambda} + S_\gamma^{n,\lambda}$ for the $n=0,\ldots, 15$ states and different values of $\lambda$.}
\end{center}
\end{figure}

A natural question that immediately arises is the overall balance between the effects of curvature on the entropy in position space and  in momentum space. As it can be seen from Figure \ref{fig:entropiessum}, for a given $\lambda$, the entropy decrease  in momentum space outweighs the raising of entropy in position space, and therefore curvature leads to a decrease of the sum of both entropies for each state. Furthermore, it can be seen that this effect increases for large values of the quantum number $n$.

\begin{figure}[H]
\begin{center}
\includegraphics[scale=0.7]{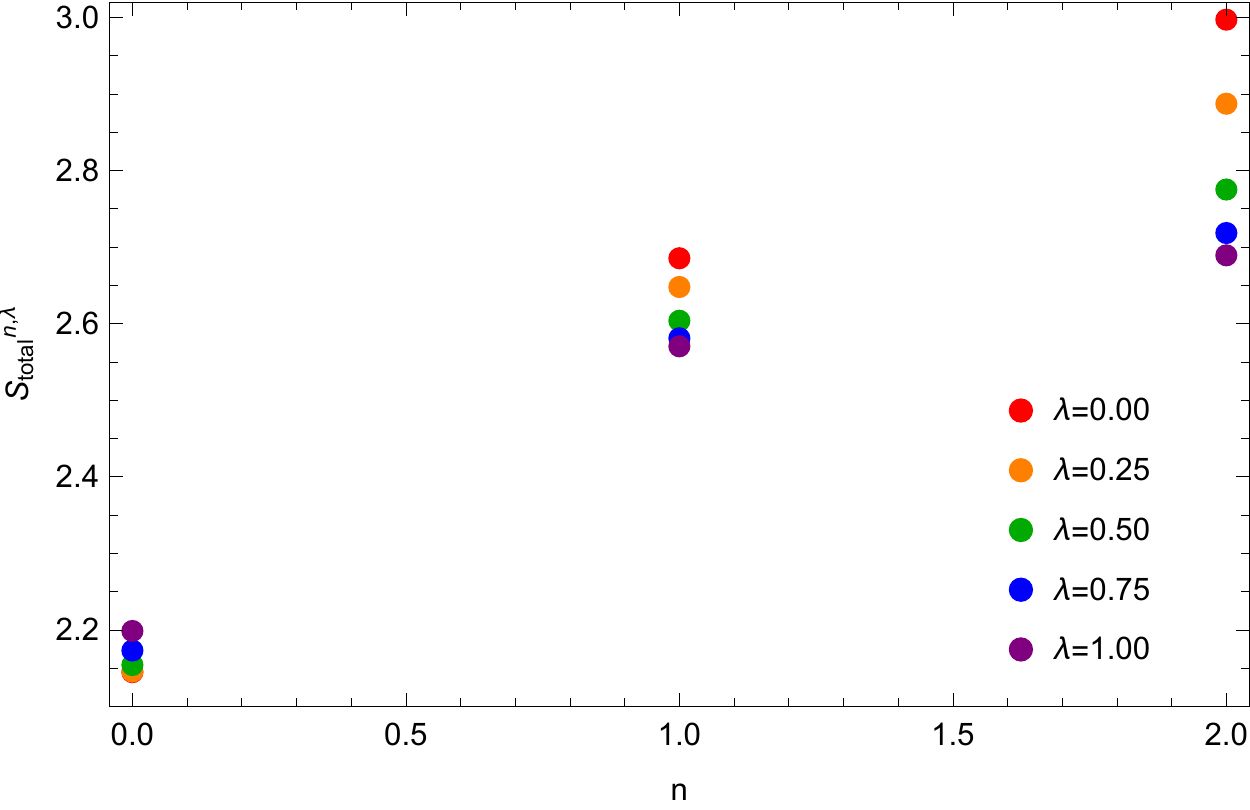}
\caption{ \label{fig:entropiessumlambdabig} 
$S_\rho^{n,\lambda} + S_\gamma^{n,\lambda}$ for the $n=0,1,2$ states and large values of $\lambda$.}
\end{center}
\end{figure}

Looking at Figure \ref{fig:entropiessum} (and its corresponding data given  in Table \ref{table:entropiessum}) it is not clear whether the ground state with $n=0$ saturates the bound \eqref{uncerSS} for any value of the curvature parameter $\lambda$. While this is obvious for $\lambda = 0$, for other values of $\lambda$ there is no reason for this to be the case. In fact, it can be numerically checked that significative differences appear when greater values of the $\lambda$ parameter are chosen, as it can be appreciated in Figure \ref{fig:entropiessumlambdabig} (which comes from numerical data given in Table \ref{table:entropiessumlambdabig}) where the total entropies for the ground, first and second excited states are plotted. We stress that, in general, smaller values of $\lambda$ are taken throughout the paper in order to guarantee the numerical accuracy of the results for larger $n$.

Figure \ref{fig:entropiessumlambdabig} and Table \ref{table:entropiessumlambdabig} show how the effect of the curvature parameter in the total entropy of the system is fundamentally different for the ground state and for the excited states. Indeed, for the ground state the total entropy grows (slowly) with $\lambda$, while for all excited states with $n\geq 1$  the total entropy decreases as a function of $\lambda$. This different behaviour can be understood by considering that the probability densities corresponding to the ground state and to the excited states have very different shapes (see Figure \ref{fig:gamma1D}),  since in the ground state the structure of maxima and zeros which is present in the excited states is completely lacking. In fact, for the ground state the probability densities in momentum space are only slightly different in terms of $\lambda$, while for the excited states the effect of the curvature becomes much more striking and localization is strongly increased. In any case, we stress that all these results are always in full agreement with the uncertainty relation \eqref{uncerSS}.


\section{Shannon information entropy in $N$ dimensions}

The aim of this section is to compute the Shannon information entropy for the N-dimensional Darboux III oscillator. Similarly to what happens in the usual harmonic oscillator, the radial symmetry leads to a wave function with radial and angular parts, and since the Darboux III oscillator is also radially symmetric, results for the angular part can be directly extracted from the previous literature.

\subsection{Eigenstates of the $N$-dimensional Darboux III oscillator}

In particular, the wave function for the eigenstates of the Darboux III oscillator written in hyperspherical coordinates can be factorized in a radial $R_{n,l}(r)$ and an angular part $Y_{{l, \bm \mu}}(\bm \theta)$, namely
\begin{equation}
\psi_{n, l, \bm \mu}^\lambda (\mathbf r) = R_{n,l}^\lambda (r) \, Y_{l, \bm \mu} (\bm{\theta}) \, ,
\label{eq:psi_r}
\end{equation}
where the radial part is given by~\cite{BEHRR2011quantum}
\begin{equation}
R_{n,l}^\lambda (r) = N_{n,l}^\lambda \sqrt{1+\lambda r^2} \; r^l \; e^{-\frac{\Omega_{n,l}^\lambda r^2}{2}} L_n^{l-1+N/2} \left( \Omega_{n,l}^\lambda r^2 \right) \, ,
\label{eq:R}
\end{equation}
with a normalization constant 
\begin{equation}
N_{n,l}^\lambda = \sqrt{\frac{2 n! (\Omega_{n,l}^\lambda)^{l+N/2}}{\Gamma (n+l+\frac{N}{2})}} \sqrt{\frac{1}{1+\left( 2n + l + \frac{N}{2}\right) \frac{\lambda}{\Omega_{n,l}^\lambda}}} \, ,
\end{equation}
where the eigenvalues and energy-dependent frequencies are given by
\begin{equation}
E_{n,l}^\lambda = - \hbar^2 \left( 2n + l + \frac{N}{2} \right)^2 + \hbar \left( 2n + l + \frac{N}{2} \right) \sqrt{\hbar^2 \lambda^2 \left( 2n + l + \frac{N}{2} \right)^2 + \omega^2} \, ,
\end{equation}
\begin{equation}
\Omega_{n,l}^\lambda := \sqrt{\omega^2 - 2 \lambda E_{n,l}^\lambda} \, .
\end{equation} 
Note that in general the $N-$dimensional wave function $\psi_{n, l, \bm \mu}^\lambda (\mathbf r)$ depends on $N$ natural numbers (quantum numbers) $\{n, \mu_1, \ldots, \mu_{N-1} \}$. To simplify the notation we write $\bm \mu = \{\mu_1, \ldots, \mu_{N-1} \}$, and moreover $\mu_1 = l$ and $\mu_{N-1} = |m|$. These quantum numbers are non-negative integer numbers constrained by the conditions $l = 0, 1, 2, \ldots$ and $\mu_{N-1} = |m| \leq \mu_{N-2} \leq \ldots \leq \mu_2 \leq \mu_1 = l$.

The angular part of the wavefunction is given by the hyperspherical harmonics (see~\cite{Vilenkin1968special,Avery1989bookhyperspherical} for a detailed description of hyperspherical harmonics and their properties), which read
\begin{equation}
Y_{{l, \bm \mu}}(\bm \theta) = M_{l,\bm \mu} e^{i \mu_{N-1} \theta_{N-1}} \prod_{k=1}^{N-2} C_{\mu_k - \mu_{k+1}}^{\alpha_k + \mu_{k+1}} (\cos \theta_k) (\sin \theta_k)^{\mu_{k+1}} \, ,
\end{equation}
where $\alpha_j = \frac{1}{2} (N-j-1)$, $C_n^{\alpha} (z)$ are the Gegenbauer polynomials ($\alpha > -\frac{1}{2}$)~\cite{NUS1991book,OMS2009atlas}, and the square of the normalization constant is given by
\begin{equation}
M_{l,\bm \mu}^{2} = \frac{1}{2 \pi} \prod_{k=1}^{N-2} \frac{\Gamma(\frac{N-k+1}{2}+\mu_{k+1}) (\mu_k-\mu_{k+1})! (\frac{N-k-1}{2} + \mu_k) (N-k+ 2 \mu_{k+1} -2)!}
{\sqrt \pi \Gamma(\frac{N-k}{2}+\mu_{k+1}) (\frac{N-k-1}{2} + \mu_{k+1}) (N-k+  \mu_k + \mu_{k+1} -2)!
} \, .
\end{equation}

\begin{figure}
\begin{center}
\includegraphics[scale=0.6]{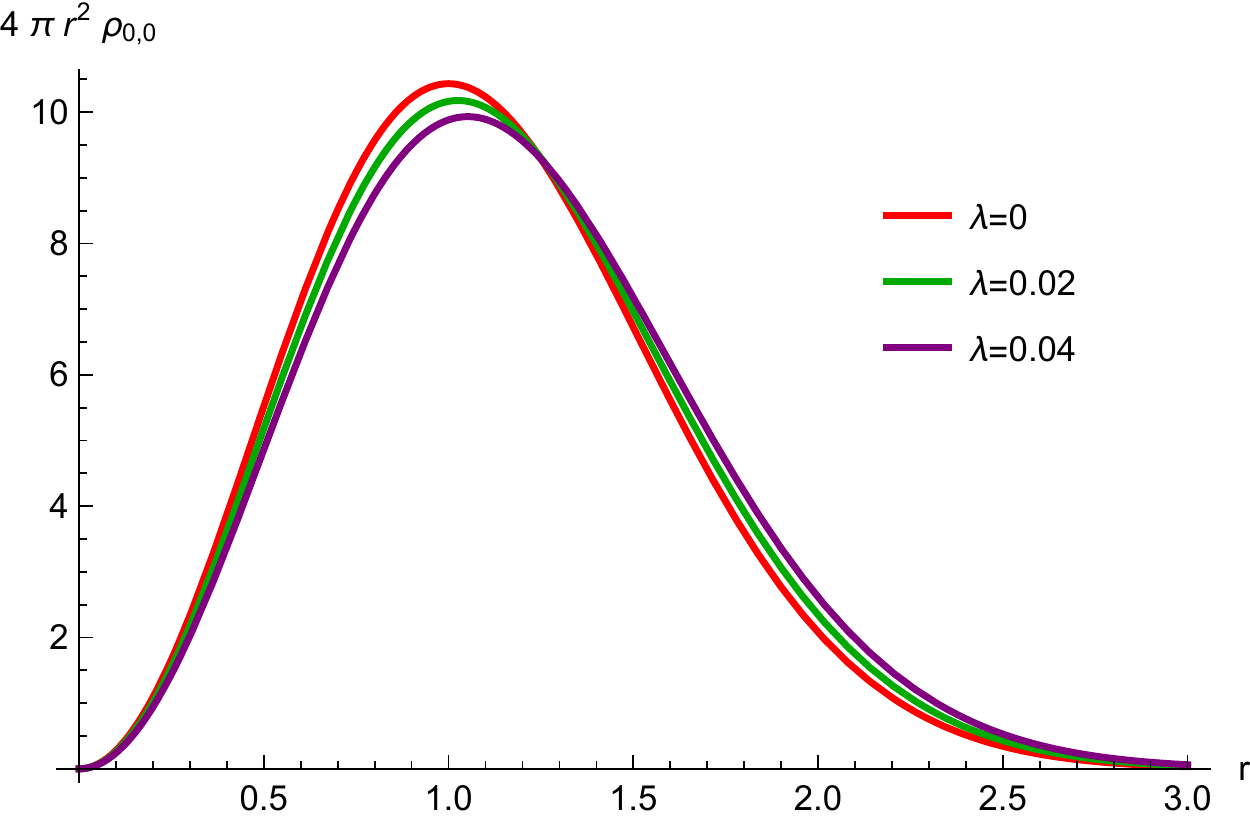}
\includegraphics[scale=0.6]{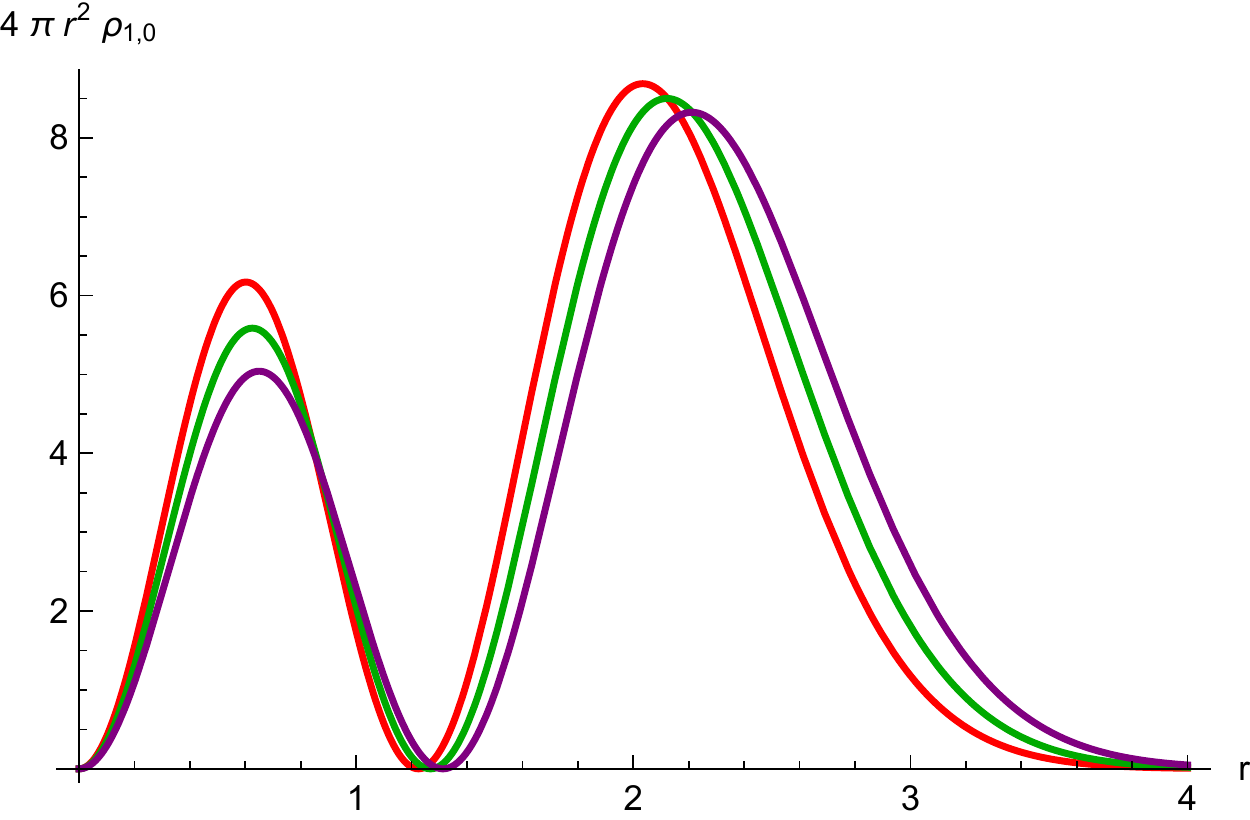} \\
\includegraphics[scale=0.6]{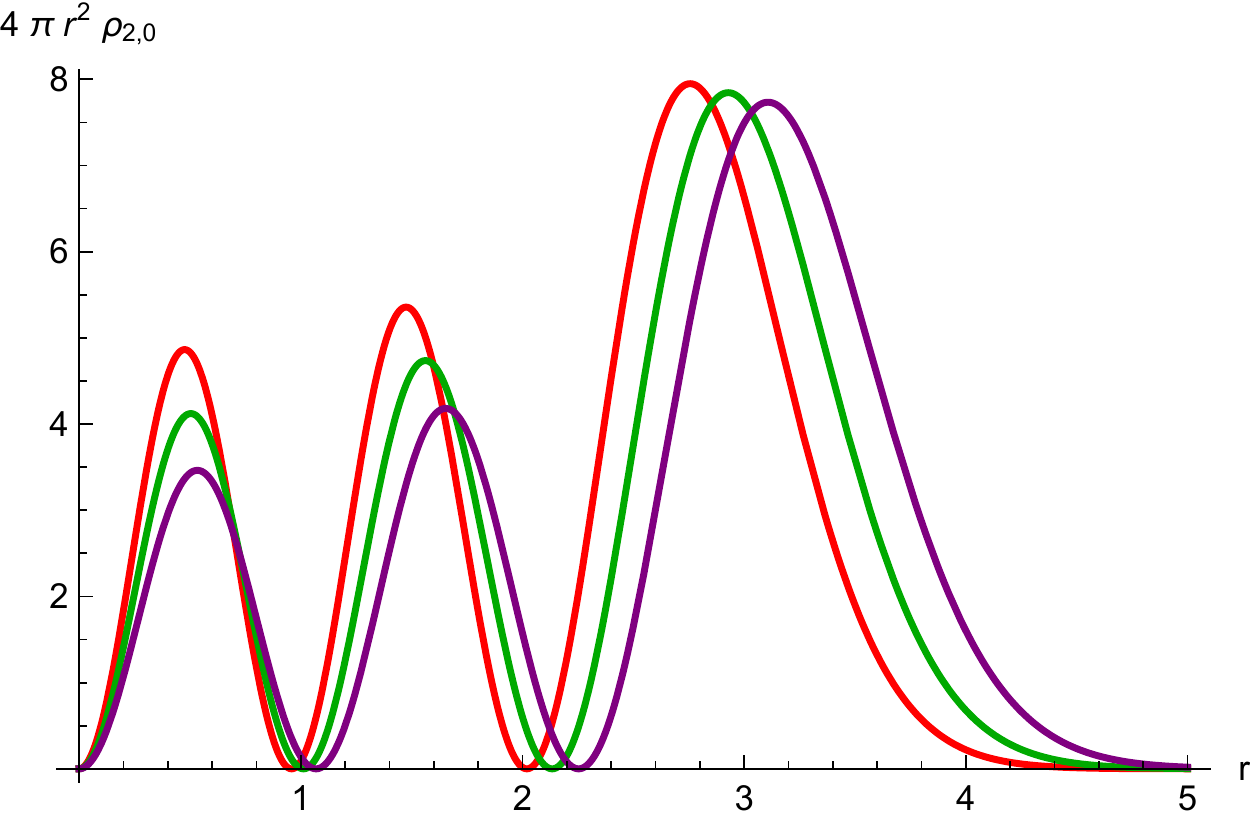} 
\includegraphics[scale=0.6]{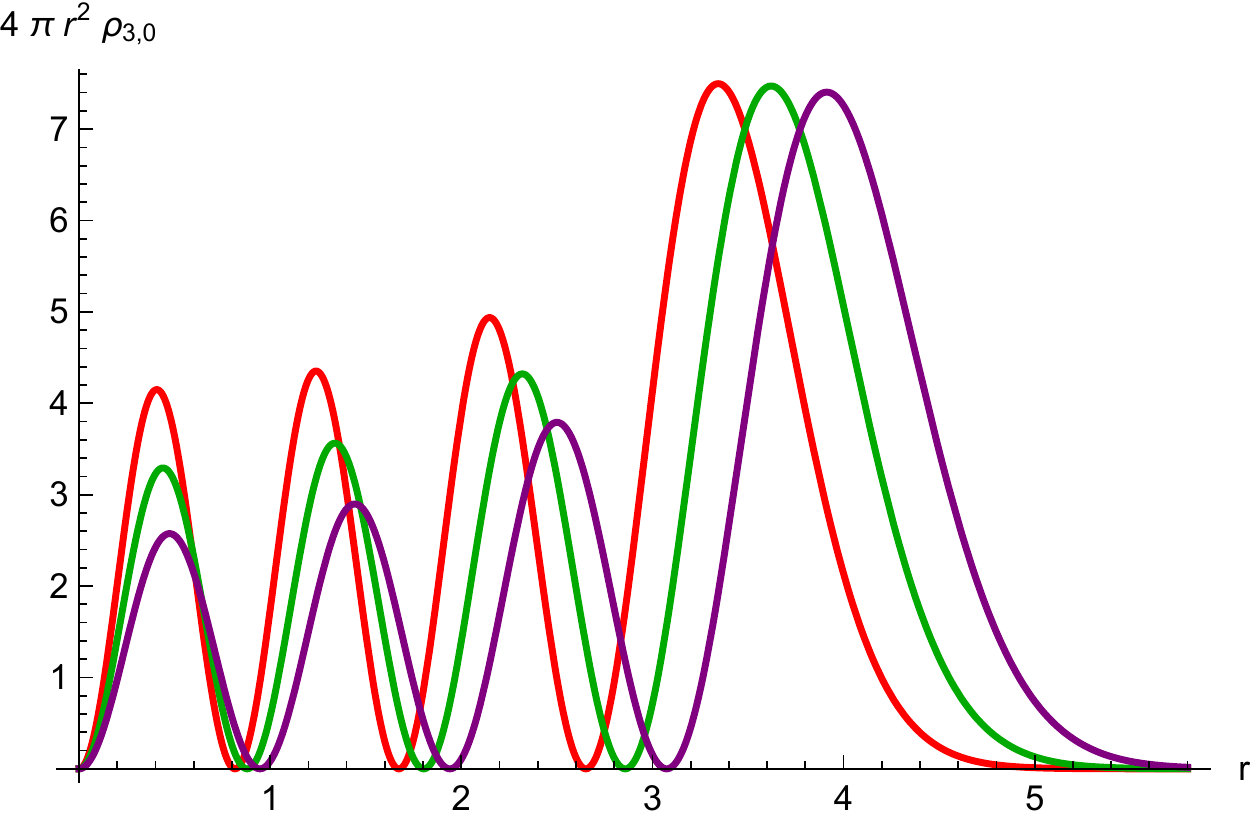}
\caption{Plots of the $N=3$ probability density of finding the particle at a distance $r$ from the origin, $4 \pi r^2 \rho_{n,0}^\lambda(r)$, for different values of $\lambda$ and different states with $n=0,1,2,3$. Top left: $n=0$. Top right: $n=1$. Bottom left: $n=2$. Bottom right: $n=3$.}
\label{fig:wafef_space3d}
\end{center}
\end{figure}

Therefore the probability density in hyperspherical coordinates reads
\begin{equation}
\begin{split}
\rho_{n, l, \bm \mu}^\lambda (\mathbf r) &= \left( R_{n,l}^\lambda (r) \right)^2 |Y_{{l, \bm \mu}}(\bm \theta)|^2 \\
&={\frac{2 n! (\Omega_{n,l}^\lambda)^{l+N/2}}{\Gamma (n+l+\frac{N}{2})}} {\frac{1}{1+\left( 2n + l + \frac{N}{2}\right) \frac{\lambda}{\Omega_{n,l}^\lambda}}} (1 + \lambda r^2) \; r^{2l} \; e^{-\Omega_{n,l}^\lambda r^2} \left( L_n^{l-1+N/2} \left( \Omega_{n,l}^\lambda r^2 \right) \right)^2
|Y_{{l, \bm \mu}}(\bm \theta)|^2 \, .
\end{split}
\end{equation}
And the radial probability density is defined as
\begin{equation}
\rho_{n, l}^\lambda (r) = \left(R_{n,l}^\lambda (r) \right)^2 \, .
\end{equation}
In the three-dimensional case the function $4 \pi r^2 \rho_{n,l}^\lambda(r)$ for the ground and first excited states by taking different values of $\lambda$ is plotted in Figure~\ref{fig:wafef_space3d}, where again the spreading influence of the curvature can be appreciated.

We also recall that the Laguerre polynomials $L_n^\alpha (z)$ are orthogonal with respect to the measure $\mathrm d \mu(z) = z^\mu e^{-z} \mathrm d z$ over $[0,+\infty)$, i.e.
\begin{equation}
\int_{0}^\infty z^\alpha L_n^\alpha (z) L_m^\alpha (z) e^{-z} \mathrm d z = 0  \, .
\end{equation}
if $m \neq n$. In the following we will need certain integrals involving Laguerre polynomials. In particular, it was shown in \cite{referee2} (see also \cite{RS1992laguerre,SMA2003laguerre}) that 
\begin{equation}
\int_{0}^\infty z^{\mu} L_n^\alpha (z) L_m^\beta (z) \, e^{-z} \mathrm d z = (-1)^{n+m} \Gamma(\mu+1)  \sum_{k=0}^{\min \{n,m\}} \binom{\mu-\alpha}{n-k} \binom{\mu-\beta}{m-k} \binom{\mu+k}{k} .
\end{equation}
Putting $m = n$ and $\beta = \alpha$ we have
\begin{equation}
\int_{0}^\infty z^{\mu} \left( L_n^\alpha (z) \right)^2 \, e^{-z} \mathrm d z = \Gamma(\mu+1) \sum_{k=0}^n \binom{\mu-\alpha}{n-k}^2 \binom{\mu+k}{k} .
\end{equation}
Fixing in this equation $\mu = \alpha$ we obtain the normalization for the Laguerre polynomials, which reads
\begin{equation}
\label{eq:Laguerrenormal}
\int_{0}^\infty z^\alpha \, \left(L_n^\alpha (z) \right)^2 \, e^{-z} \mathrm d z = \frac{\Gamma (n + b +1)}{n!}  \, ,
\end{equation}
for all $n \in \mathbb N$. Moreover, we will be interested in the special cases when $\mu = \alpha+1$: 
\begin{equation}
\label{eq:alpha+1}
\int_{0}^\infty z^{\alpha+1} \left( L_n^\alpha (z) \right)^2 \, e^{-z} \mathrm d z = \frac{\Gamma (n + \alpha +1)}{n!} (2n+\alpha+1) \, ,
\end{equation}
and $\mu = \alpha+2$: 
\begin{equation}
\label{eq:alpha+2}
\int_{0}^\infty z^{\alpha+2} \left( L_n^\alpha (z) \right)^2 \, e^{-z} \mathrm d z = \frac{\Gamma (n + \alpha +1)}{n!} (6n^2 + 6 n + 6 n \alpha + 3 \alpha + \alpha^2 + 2) \, .
\end{equation}

\subsection{Shannon entropy in arbitrary dimension}

With all the previous results at hand, the Shannon information entropy for the $N$-dimensional Darboux III oscillator can be computed. The factorization of the wave function in its radial and angular parts leads to 
\begin{equation}
\begin{split}
&S_\rho^{N,n,l,\bm \mu, \lambda} = - \int_{\mathbb R^N} \rho_{n, l, \bm \mu}^\lambda (\mathbf r) \log \bigg( \rho_{n, l, \bm \mu}^\lambda (\mathbf r) \bigg) \mathrm d \mathbf r = \\
&- \int_{\mathbb R^N} \left( R_{n,l}^\lambda (r) \right)^2 |Y_{{l, \bm \mu}}(\bm \theta)|^2 \log \bigg( \left( R_{n,l}^\lambda (r) \right)^2 \bigg) \mathrm d \mathbf r -\int_{\mathbb R^N} \left( R_{n,l}^\lambda (r) \right)^2 |Y_{{l, \bm \mu}}(\bm \theta)|^2 \log \bigg( |Y_{{l, \bm \mu}}(\bm \theta)|^2 \bigg) \mathrm d \mathbf r  \\
&= - \int_0^\infty \left( r R_{n,l}^\lambda (r) \right)^2 \log \bigg( \left( R_{n,l}^\lambda (r) \right)^2 \bigg) \mathrm d r \int_{\mathbb S^{N-1}} |Y_{{l, \bm \mu}}(\bm \theta)|^2 \mathrm d \Omega_N \\
&- \int_0^\infty \left( r R_{n,l}^\lambda (r) \right)^2 \mathrm d r \int_{\mathbb S^{N-1}} |Y_{{l, \bm \mu}}(\bm \theta)|^2 \log \bigg( |Y_{{l, \bm \mu}}(\bm \theta)|^2 \bigg) \mathrm d \Omega_N \\
&=- \int_0^\infty \left( r R_{n,l}^\lambda (r) \right)^2 \log \bigg( \left( R_{n,l}^\lambda (r) \right)^2 \bigg) \mathrm d r - J_Y ,
\end{split} 
\label{eq:S_NDcomp}
\end{equation}
where 
\begin{equation}
J_Y := \int_{\mathbb S^{N-1}} |Y_{{l, \bm \mu}}(\bm \theta)|^2 \log \bigg( |Y_{{l, \bm \mu}}(\bm \theta)|^2 \bigg) \mathrm d \Omega_N .
\end{equation}
Using the integrals \eqref{eq:Laguerrenormal}, \eqref{eq:alpha+1} and \eqref{eq:alpha+2}, the integral involving the radial part of the wave function can be expressed completely in terms of integrals involving Laguerre polynomials.  In this way, the Shannon entropy on position space of an arbitrary quantum state of the $N$-dimensional Darboux III oscillator is given by
\begin{equation}
\label{eq:S_ND}
\begin{split}
S_\rho^{N,n,l,\bm \mu, \lambda} &= - \frac{N}{2} \log \Omega_{n,l}^\lambda - \log \bigg( \frac{2 n!}{\Gamma(n+l+ \frac{N}{2})} \bigg) + \log \bigg( 1+\bigg( 2n+l+\frac{N}{2} \bigg) \frac{\lambda}{\Omega_{n,l}^\lambda} \bigg) \\
&- \frac{n!}{\Gamma(n+l+ \frac{N}{2})} \frac{1}{1+\big( 2n+l+\frac{N}{2} \big) \frac{\lambda}{\Omega_{n,l}^\lambda}} \bigg( J_1 + \frac{\lambda}{\Omega_{n,l}^\lambda} \tilde J_1 + J_2 + \frac{\lambda}{\Omega_{n,l}^\lambda} \tilde J_2 + J_3^{\frac{\lambda}{\Omega_{n,l}^\lambda}} + \frac{\lambda}{\Omega_{n,l}^\lambda} \tilde J_3^{\frac{\lambda}{\Omega_{n,l}^\lambda}} \bigg) \\
&+ \frac{1}{1+\big( 2n+l+\frac{N}{2} \big) \frac{\lambda}{\Omega_{n,l}^\lambda}} \bigg( \big( 2n+l+\frac{N}{2} \big) + \frac{\lambda}{\Omega_{n,l}^\lambda} \big( 6 \, n^2 + 6 \, l \, n + 3 \, n \, N + l + l^2 + l \, N + \frac{N}{2} + \frac{N^2}{4}\big)\bigg) - J_Y \, ,
\end{split} 
\end{equation}
where 
\begin{equation}
J_1 = \int_0^{+\infty} z^{l+\frac{N}{2}-1} e^{-z} \bigg( L_n^{l+\frac{N}{2}-1} (z) \bigg)^2 \log(z^l) \, \mathrm d z \, ,
\end{equation}

\begin{equation}
\tilde J_1 = \int_0^{+\infty} z^{l+\frac{N}{2}} e^{-z} \bigg( L_n^{l+\frac{N}{2}-1} (z) \bigg)^2 \log(z^l) \, \mathrm d z \, ,
\end{equation}

\begin{equation}
J_2 = \int_0^{+\infty} z^{l+\frac{N}{2}-1} e^{-z} \bigg( L_n^{l+\frac{N}{2}-1} (z) \bigg)^2 \log \bigg( \bigg( L_n^{l+\frac{N}{2}-1} (z) \bigg)^2 \bigg) \, \mathrm d z \, ,
\end{equation}

\begin{equation}
\tilde J_2 = \int_0^{+\infty} z^{l+\frac{N}{2}} e^{-z} \bigg( L_n^{l+\frac{N}{2}-1} (z) \bigg)^2 \log \bigg( \bigg( L_n^{l+\frac{N}{2}-1} (z) \bigg)^2 \bigg) \, \mathrm d z \, ,
\end{equation}

\begin{equation}
J_3^\alpha = \int_0^{+\infty} z^{l+\frac{N}{2}-1} e^{-z} \bigg( L_n^{l+\frac{N}{2}-1} (z) \bigg)^2 \log \bigg( 1 + \alpha \, z \bigg) \, \mathrm d z \, ,
\end{equation}

\begin{equation}
\tilde J_3^\alpha = \int_0^{+\infty} z^{l+\frac{N}{2}} e^{-z} \bigg( L_n^{l+\frac{N}{2}-1} (z) \bigg)^2 \log \bigg( 1 + \alpha \, z \bigg) \, \mathrm d z \, .
\end{equation}

The well-known Shannon entropy for the harmonic oscillator is recovered in the limit $\lambda \to 0$,
\begin{equation}
S_\rho^{N,n,l,\bm \mu, 0} := \lim_{\lambda \to 0} S_\rho^{N,n,l,\bm \mu, \lambda} \, ,
\end{equation}
thus obtaining 
\begin{equation}
S_\rho^{N,n,l,\bm \mu, 0} = - \frac{N}{2} \log \omega - \log \bigg( \frac{2 n!}{\Gamma(n+l+ \frac{N}{2})} \bigg) - \frac{n!}{\Gamma(n+l+ \frac{N}{2})} (J_1 + J_2) + 2 \, n + l + \frac{N}{2} \, ,
\end{equation}
which is exactly the same expression given in \cite{DAY1997entropy} since $J_1$ and $J_2$ are independent of the curvature $\lambda$.

\subsection{The three-dimensional case}

Since the angular part of the wave functions defining the Darboux III eigenstates is the same as in the harmonic oscillator, the integral $J_Y$ involving the spherical harmonics is also exactly the same. For the particular case $N=3$, it turns out that $J_Y$ can be expressed in terms of the Gegenbauer polynomials, as it was shown in \cite{DAY1997entropy}. For the sake of completeness we reproduce this explicit result here and for $N=3$ we have that
\begin{equation}
\begin{split}
J_Y = \log \bigg( \frac{(2l+1)(l-m)!}{4 \pi (l+m)!} \bigg) + \log \bigg( \frac{(2l+1)(l-m)! \big( (2m)! \big)^2}{2^{2m+1}  (l+m)! (m!)^2} \bigg) (J_{Y,1} +J_{Y,2}) + 2 \log \bigg( \frac{(2m)!}{m! 2^m} \bigg) ,
\end{split}
\end{equation}
where
\begin{equation}
J_{Y,1} = \int_{-1}^1 \bigg( C_{l-m}^{m+\frac{1}{2}} (z) \bigg)^2 (1-z^2)^m \log (1-z^2)^m \mathrm d z ,
\end{equation}
and
\begin{equation}
J_{Y,2} = \int_{-1}^1 \bigg( C_{l-m}^{m+\frac{1}{2}} (z) \bigg)^2 (1-z^2)^m \log \bigg( C_{l-m}^{m+\frac{1}{2}} (z) \bigg)^2 \mathrm d z .
\end{equation}
for $m \geq 0$. If $m<0$ the previous formulas are still valid substituting $m$ by $-m$.

\begin{figure}[t]
\begin{center}
\includegraphics[scale=0.6]{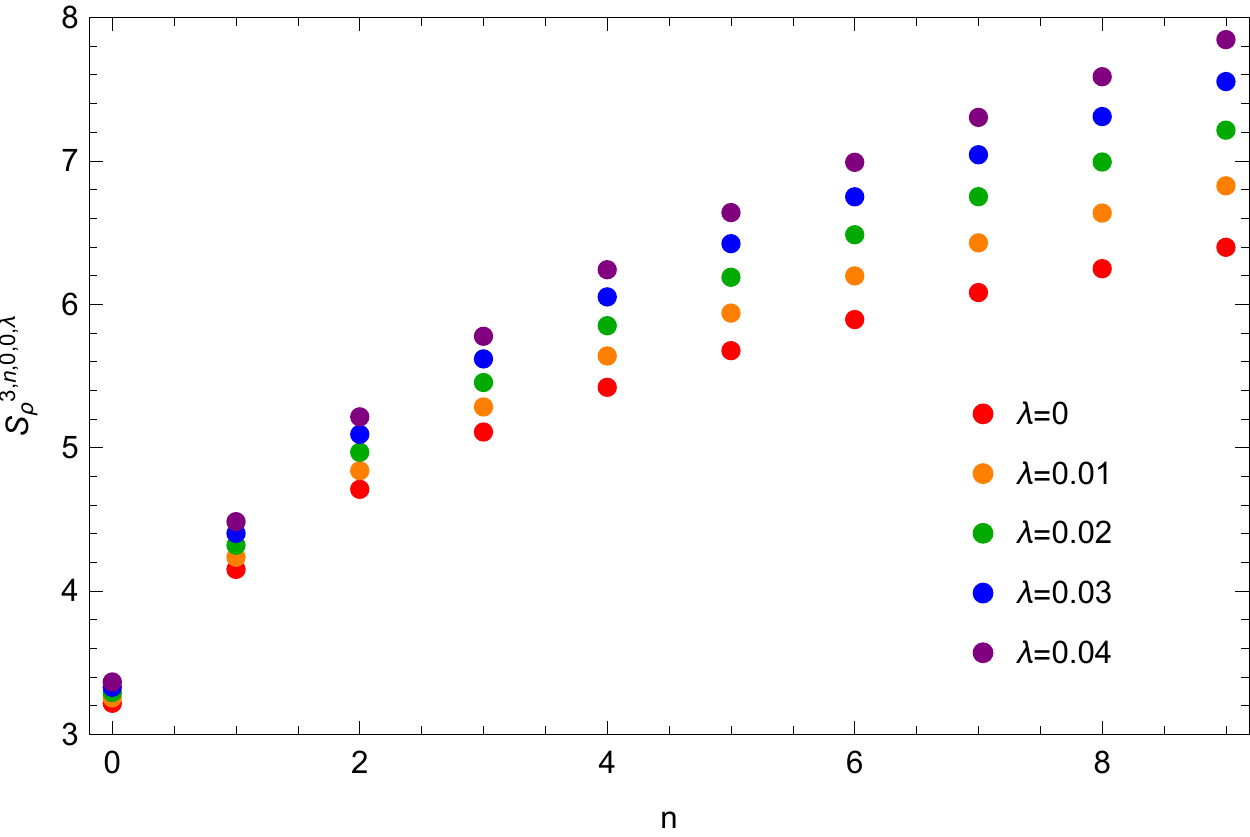}
\caption{\label{fig:entropiesspace3d}
Entropies $S_\rho^{3,n,0,0, \lambda}$ for the states with $n=0,\ldots, 9$ and different values of $\lambda$.}
\end{center}
\end{figure}

For this three-dimensional case, Figure~\ref{fig:entropiesspace3d} contains the plots of the entropies $S_\rho^{N,n,l,\bm \mu, \lambda}$ for the ground and some low excited states of the Darboux III oscillator with vanishing $l$ and $m$ quantum numbers {(see Table \ref{table:entropiesspace3d} for the numerical data)}. As a general pattern, the entropy increases with $\lambda$ for fixed $n$. On the other hand, and similarly to what happens in the non-deformed harmonic oscillator with $\lambda=0$, it can be numerically checked that for fixed $\lambda$ and $n$ the entropy in position space grows with $l$, and for a given $\lambda$ and $l$ the entropy increases with $n$. Therefore, the curvature increases the spreading of the wave function in position space, without altering the qualitative effects due to $n$ and $l$ which are already present in the flat three-dimensional harmonic oscillator states.

\subsection{Shannon entropy in momentum space for $N=3$}

\begin{figure}[H]
\begin{center}
\includegraphics[scale=0.6]{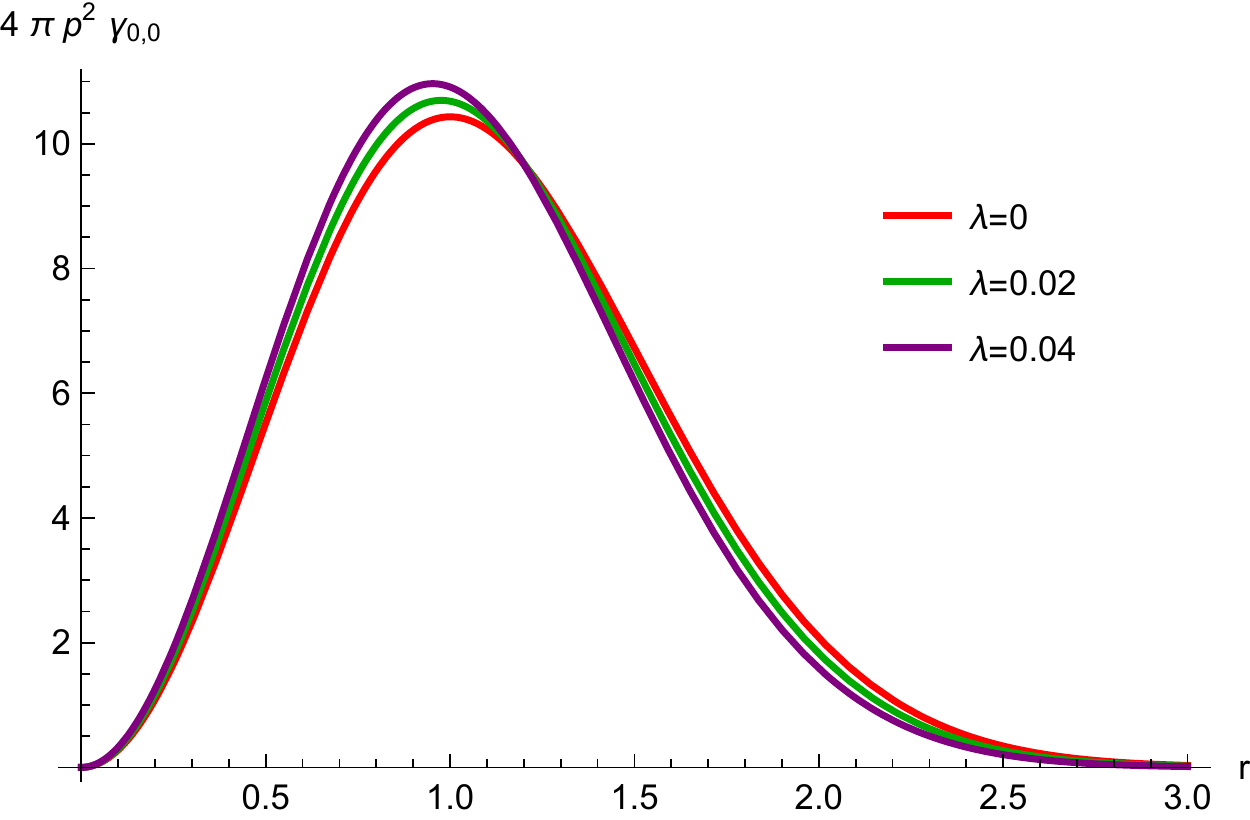}
\includegraphics[scale=0.6]{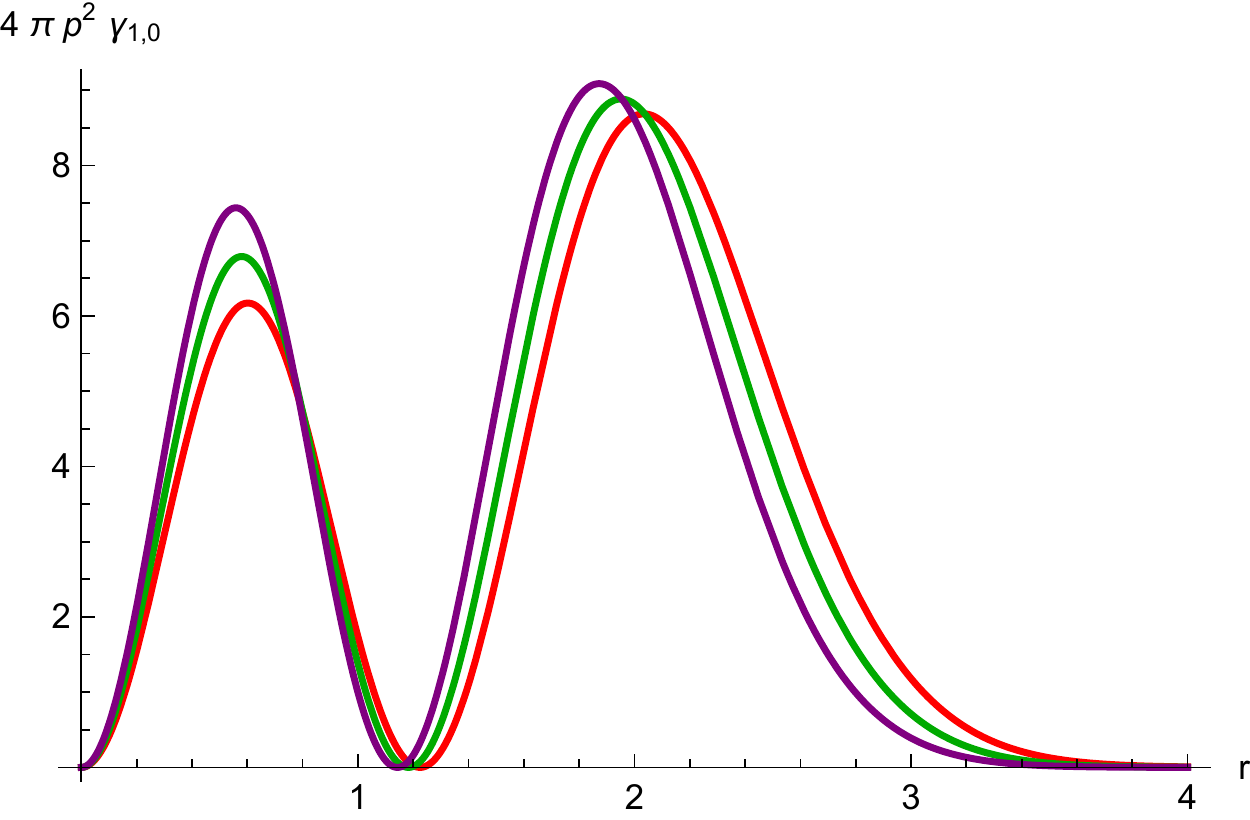} \\
\includegraphics[scale=0.6]{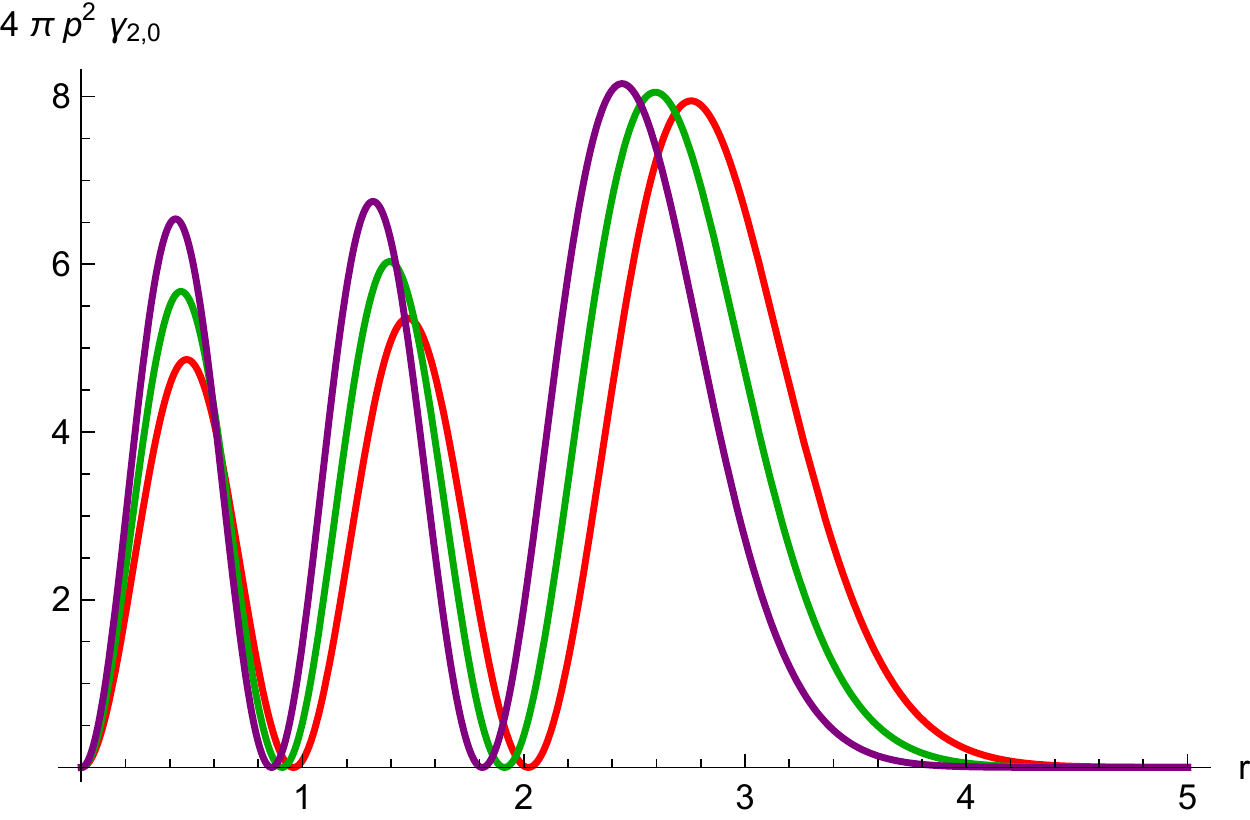} 
\includegraphics[scale=0.6]{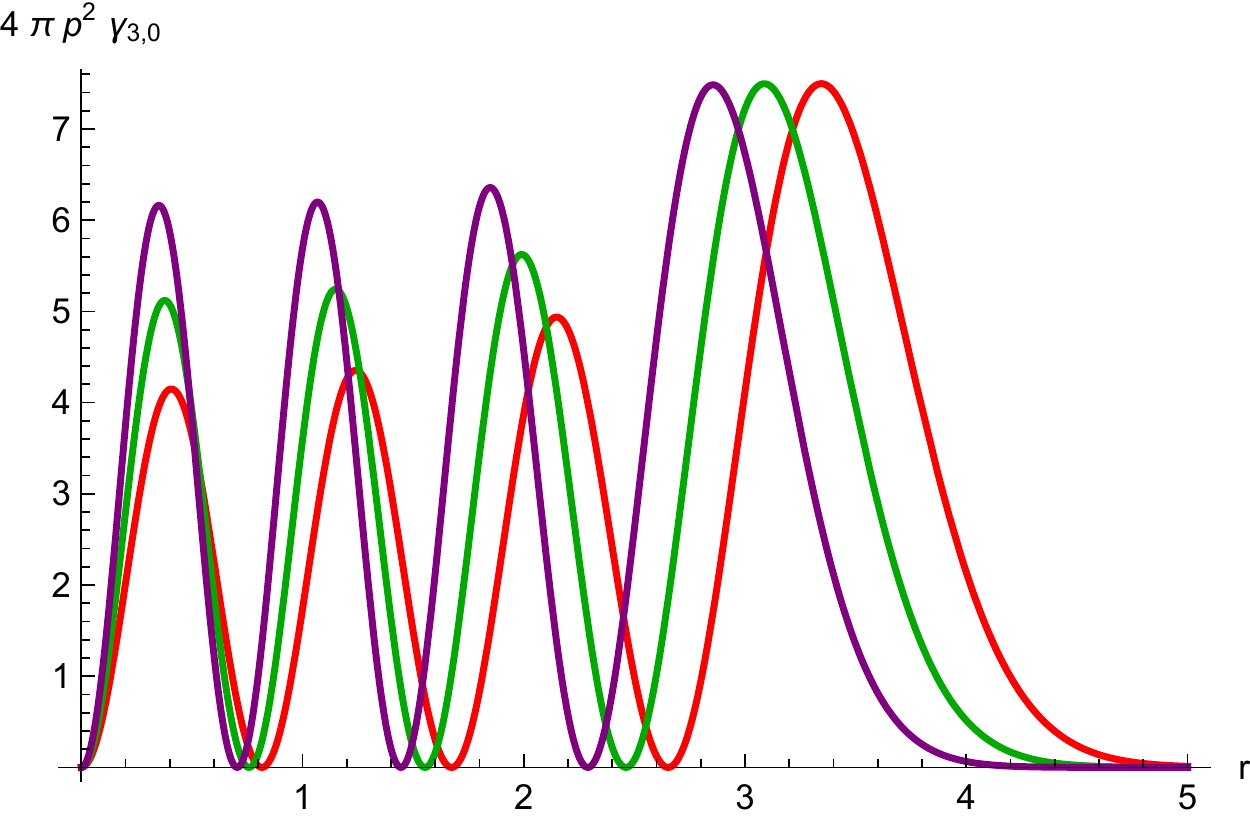}
\caption{Plots of the $N=3$ probability density $4 \pi p^2 \gamma_{n,0}^\lambda(r)$ of finding the particle with a radial momentum $p$, for different values of $\lambda$ and different states with $n=0,1,2,3$ and  $l=0$. Top left: $n=0$. Top right: $n=1$. Bottom left: $n=2$. Bottom right: $n=3$.}
\label{fig:wafef_momentum3d}
\end{center}
\end{figure}

Similarly to what happens in the one-dimensional case, the non-constant curvature of the Darboux III space prevents analytical expressions for the entropies in momentum space. Nevertheless, the radial symmetry of the system makes it possible to perform a numerical analysis of the behaviour of said entropies (see~\cite{Olendski} for a similar approach applied to spherical quantum dots). In particular, in order to compute for $N=3$ the Fourier transform of \eqref{eq:psi_r}, 
namely
\begin{equation}
\label{eq:psi3D}
\tilde \psi_{n, l, m}^\lambda (\mathbf p) = \mathcal F \left\{ \psi_{n, l, m}^\lambda (\mathbf r) \right\} (\mathbf p) = \left( \frac{1}{2 \pi} \right)^{3/2} \int_{\mathbb R^3} e^{i \mathbf p \cdot \mathbf r} \psi_{n, l, m}^\lambda (\mathbf r) \,  \mathrm d \mathbf r ,
\end{equation}
we use the well known \cite{Avery1989bookhyperspherical} spherical expansion of plane waves given by
\begin{equation}
e^{i \mathbf p \cdot \mathbf r} = 4 \pi \sum_{l'} i^{l'} j_{l'} (p \, r) \sum_{m'} Y_{l' m'} (\bm{\theta_{p}}) Y_{l' m'}^* (\bm{\theta_{r}}) ,
\label{eq:expFourier}
\end{equation}
where $j_{l} (z)$ stands for the spherical Bessel function, which is related to ordinary Bessel function by $j_{l} (z) := \sqrt{\frac{\pi}{2}} z^{-1/2} J_{l+\frac{1}{2}} (z)$. Moreover, in the previous expression \eqref{eq:expFourier} we have introduced the notation $\bm{\theta_{r}}$ and $\bm{\theta_{p}}$ to denote the angular variables in position and momentum space, respectively. Introducing \eqref{eq:expFourier} into \eqref{eq:psi3D} we obtain a similar expression to \eqref{eq:psi_r} in momentum space
\begin{equation}
\label{eq:psi_k_3D}
\tilde \psi_{n, l, m}^\lambda (\mathbf p) = i^l \mathcal K_{n,l}^\lambda (p) Y_{l m} (\bm{\theta_{p}}) ,
\end{equation}
where
\begin{equation}
\mathcal K_{n,l}^\lambda (p) := \sqrt{\frac{2}{\pi}} \int j_l (p \, r) R_{n,l}^\lambda (r) r^2 \mathrm d r .
\end{equation}
In the particular case $N=3$, the function $R_{n,l}^\lambda (r)$ is given by~\eqref{eq:R}. Using this factorization of the wavefunction in momentum space, all of our previous results (namely \eqref{eq:S_NDcomp} and below) could be easily applied if $\mathcal K_{n,l}^\lambda (p)$ would admit an analytical expression but, similarly to the one-dimensional situation, this is not the case. Nevertheless, it is possible to perform all the numerical computations needed in order to obtain the radial probability density in momentum space given by
\begin{equation}
\gamma_{n, l}^\lambda (p) = \left(\mathcal K_{n,l}^\lambda (p) \right)^2 \, .
\end{equation}
In fact, Figure \ref{fig:wafef_momentum3d} shows the function $4 \pi r^2 \gamma_{n,l}^\lambda(p)$ for the $n=0,1,2,3$ states and by taking $l=0$, where the influence of the curvature in the compression of the radial probability density in momentum space is clearly appreciated (note that, again, this effect is much less significative for the ground state).

\begin{figure}[H]
\begin{center}
\includegraphics[scale=0.7]{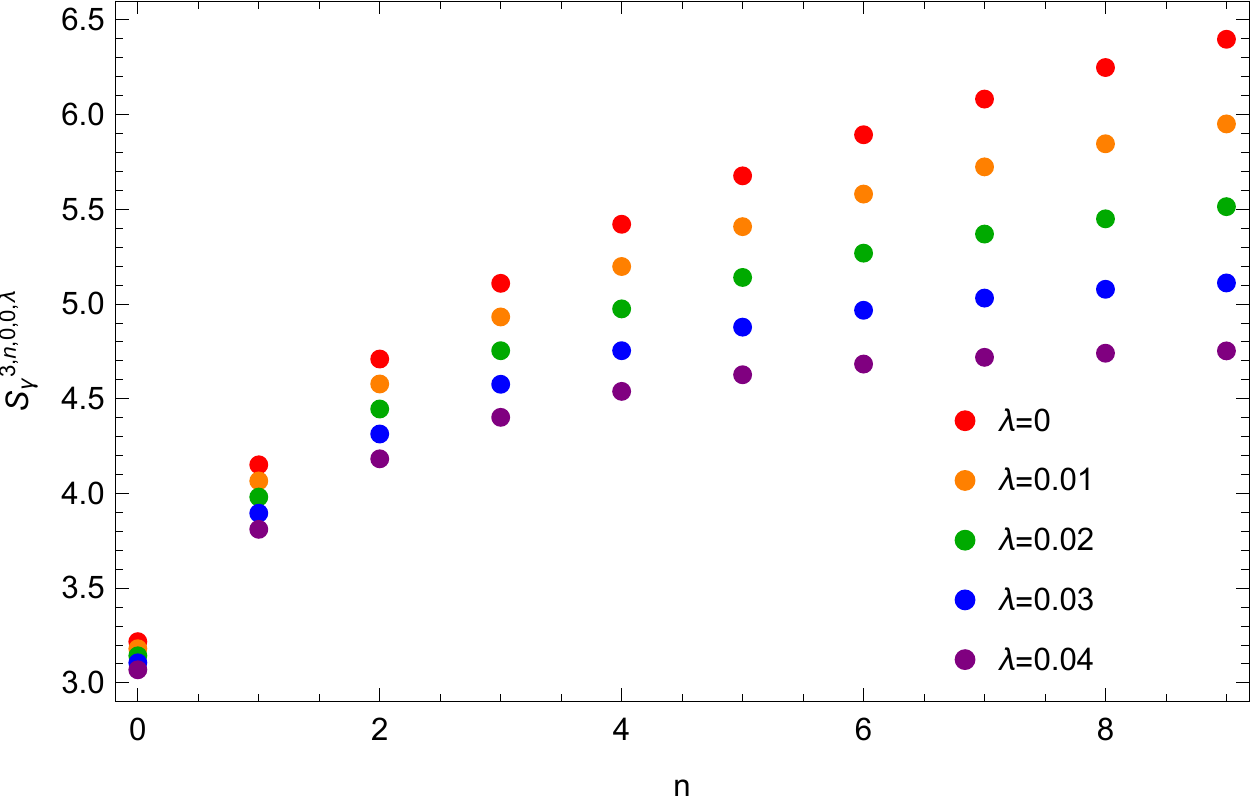}
\caption{\label{fig:entropiesmomentum3D}
Entropies $S_\gamma^{3,n,0,0, \lambda}$ for the states with $n=0,\ldots, 9$ and different values of $\lambda$.}
\end{center}
\end{figure}

In this manner, any of the entropies $S_\gamma^{3,n,l,m, \lambda}$ could be computed numerically. In particular, Figure \ref{fig:entropiesmomentum3D} (with data shown in Table \ref{table:entropiesmomentum3D}) plots the entropies in momentum space for the states $n=0,\ldots, 9$ with $l=m=0$ and different values of $\lambda$. As it can be easily seen by comparing this plot with Figure \ref{fig:entropiesmomentum}, the trend of the dependence of $S_\gamma$ for $N=3$ in terms of the curvature parameter $\lambda$ is exactly the same as in the one-dimensional system: for a given $\lambda$ the entropy $S_\gamma$ grows with $n$ and for a fixed $n$ it decreases with $\lambda$. 

\begin{figure}[H]
\begin{center}
\includegraphics[scale=1]{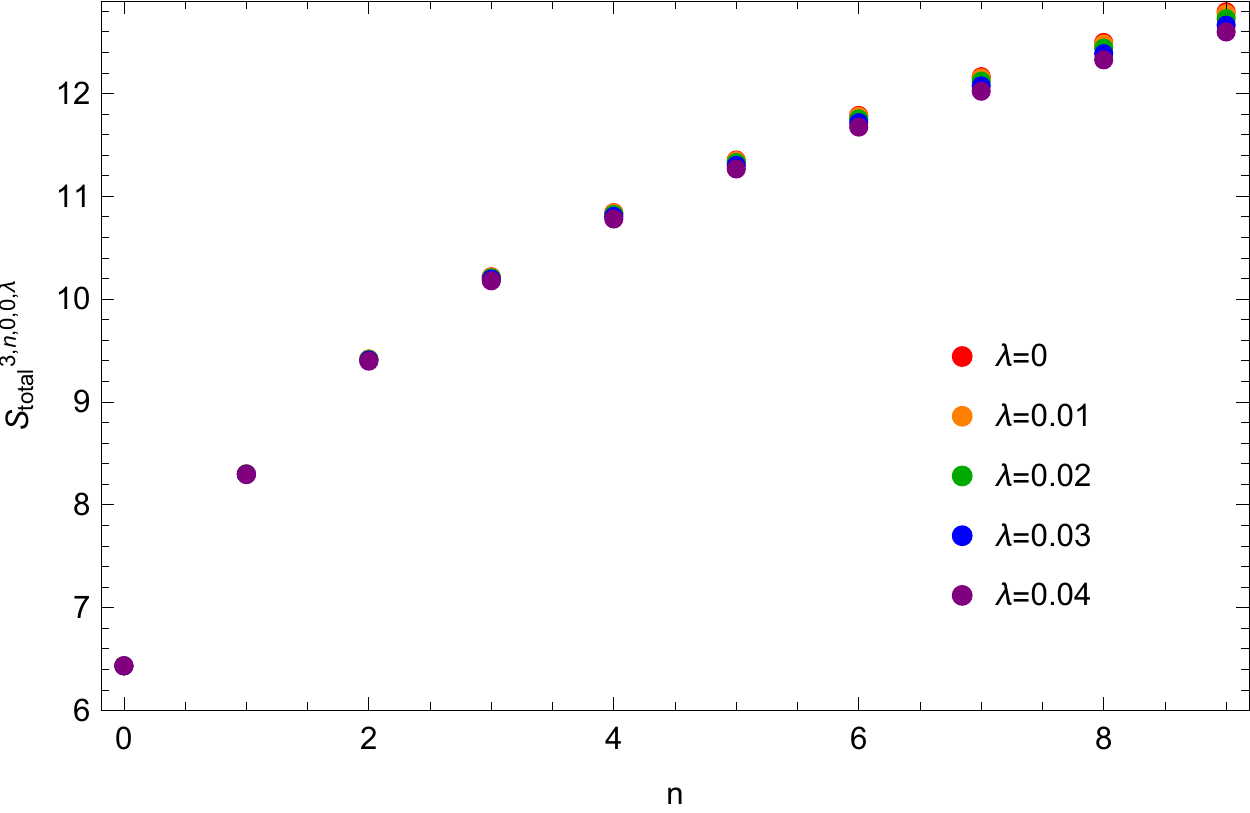}
\includegraphics[scale=0.6]{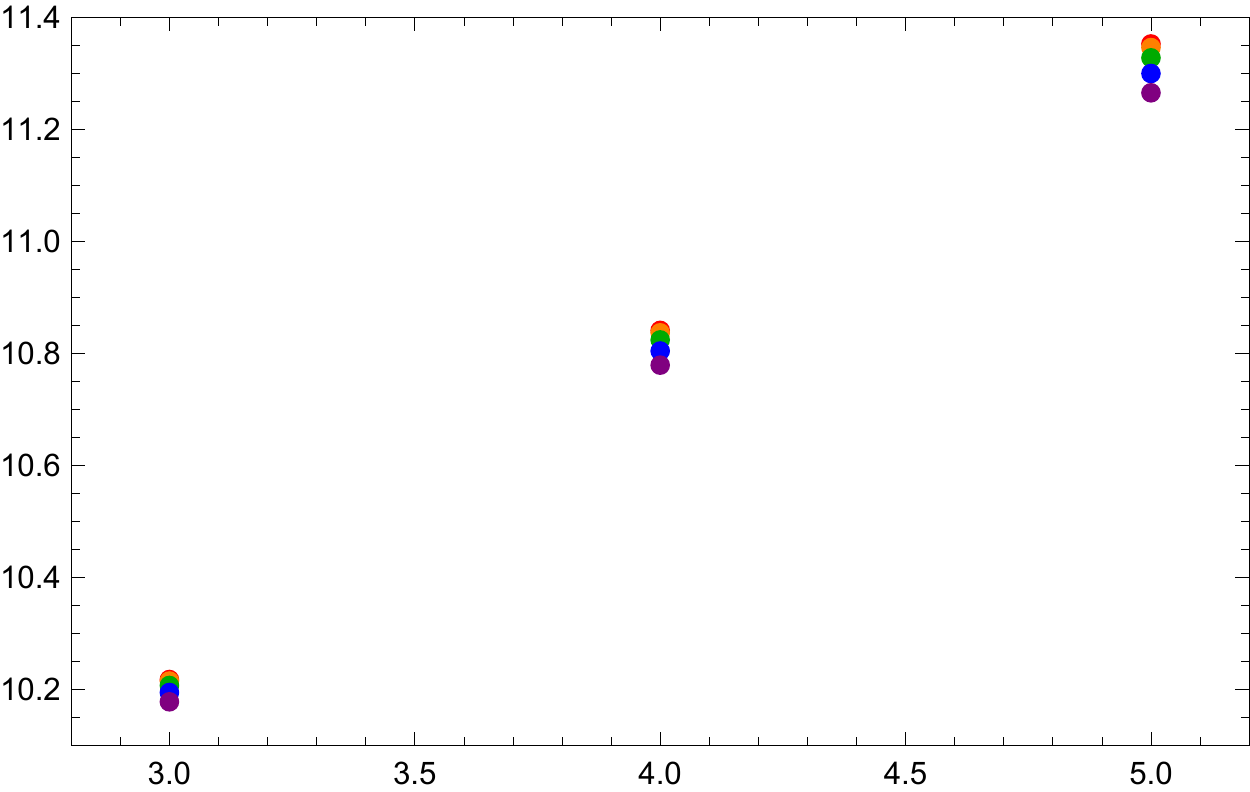}\hspace{2cm}\includegraphics[scale=0.6]{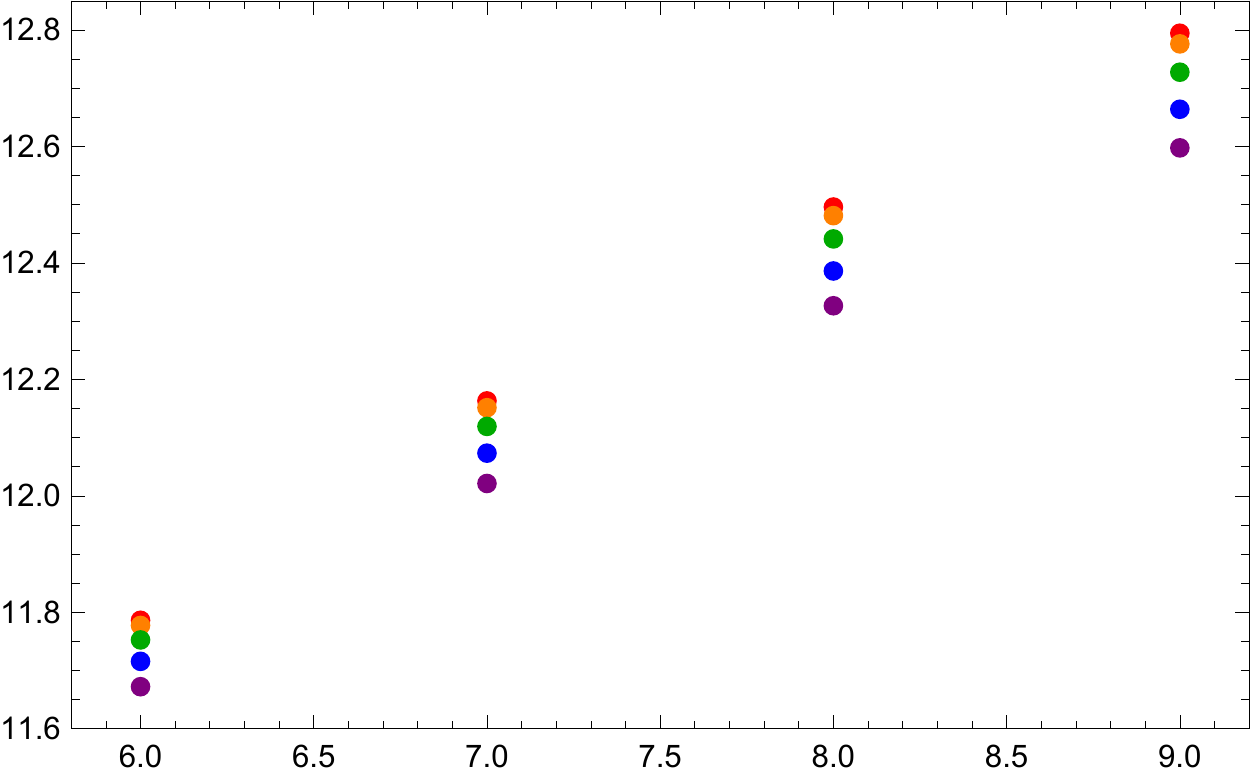}
\caption{\label{fig:entropiessum3d}
Entropies $S_{\mathrm{total}}^{3,n,0,0, \lambda}$ for the states with $n=0,1,\dots,9$ and different values of the curvature $\lambda$. The two boxes below are enlarged fragments of the plot where the variation of the entropy in terms of $\lambda$ can be more easily appreciated.}
\end{center}
\end{figure}

Finally, in Figure \ref{fig:entropiessum3d} (data in Table \ref{table:entropiessum3d}) the total entropies for the very same three-dimensional Darboux III states are presented. We stress that their qualitative behaviour coincides with the one-dimensional system plotted in Figure \ref{fig:entropiessum}. We remark that, as it happened in Figure \ref{fig:entropiessum}, for the chosen values of $\lambda$ and for the numerical precision in Table \ref{table:entropiessum3d}, in the ground state $n=0$ all the total entropies seem to coincide. However, it can be checked numerically that this is no longer the case for larger values of $\lambda$ and, once again, only in the ground state the total entropy grows with $\lambda$ while for any excited state it becomes smaller as the curvature parameter grows. Therefore, for the three-dimensional system we obtain an analogue of Figure \ref{fig:entropiessumlambdabig}.

\section{Concluding remarks}

In this paper we have computed the Shannon information entropy $S_\rho$ for the position representation of an arbitrary eigenstate of a $N$-dimensional nonlinear quantum oscillator defined on a space whose negative non-constant curvature is governed by a non-negative parameter $\lambda$, and we have expressed $S_\rho$ completely in terms of integrals involving orthogonal polynomials. Our results indeed reproduce the known ones for the $N$-dimensional harmonic oscillator when the curvature of the space vanishes, {\em i.e.} in the $\lambda\to 0$ limit. The computation on position space has been performed analytically in arbitrary dimension (see \eqref{eq:entropy1Dspace} for the one-dimensional case and \eqref{eq:S_ND} for dimension $N>1$). Moreover, computations of the Shannon entropy in momentum space were performed numerically in the one and three-dimensional cases, since the modification induced by the curvature in the wave functions of the Darboux III oscillator makes it impossible to obtain an analytical expression for the Fourier transformed eigenstates.

We have found that in the $N$-dimensional case the effect of a larger absolute value of the negative curvature (through a bigger $\lambda$) is to increase the entropy in position space. The effect on wavefunctions defined on momentum space is found to be the opposite one, and for all excited states with quantum number $n\geq 1$ the decrement of the entropy in momentum space outweighs the increment on position space. Therefore, this work can be thought of as a first step in the study of the interplay between curvature in quantum mechanical systems and their information theoretic properties.

Within this perspective, it is worth stressing that the Darboux III oscillator is an exactly solvable nonlinear quantum model that presents two remarkably interesting features: firstly, is analytically solvable in any dimension and, secondly, it includes an additional parameter $\lambda$, which carries a geometrical interpretation and whose smooth $\lambda\to 0$ limit leads to the well-known results for the harmonic oscillator. Moreover, the exact solvability of the Darboux III quantum oscillator is a structural property of this nonlinear oscillator which holds for any value of $\lambda$. This means that all the information-theoretic problems that have been previously studied in the literature for the usual $N$-dimensional oscillator can be also faced for the Darboux III oscillator here presented, and the role of the curvature within them can be always analysed through the parameter $\lambda$. 

Among these problems, we can quote the analysis of the $N$-dimensional Darboux III oscillator in terms of cartesian coordinates~\cite{ToranzoDehesa2019} (instead of the hyperspherical ones here used), the computation of its Renyi and Tsallis information entropies as well as its radial expectation values~\cite{ATTD2016,Dehesa2017EntropicMO,Puertas-Centeno:2018yin}, the study of the Rydberg states obtained for large values of the quantum number $n$~\cite{VAYD95,MO96,JSR97,Dehesa2017EntropicMO,ToranzoDehesaRydberg} and also the properties of the large $N$ limit~\cite{Puertas-Centeno:2017jpv,Dehesa:2019nhu,ToranzoDehesaRydberg} of the new nonlinear oscillator model here presented. 
Finally, the question concerning the possibility of the analytical description of information-theoretic properties for $N$-dimensional hydrogenic systems on curved spaces raises in a natural way. The answer is affirmative, since in~\cite{Ballesteros20091219,BEHRR2014Taub} an exactly solvable (and also maximally superintegrable) deformation of the Coulomb problem on a space with non-constant curvature was presented.  The corresponding Hamiltonian is given by the sum of the kinetic energy corresponding to a $N$-dimensional radially symmetric space which is related to the well-known Taub-NUT space (see~\cite{BEHRR2014Taub} and references therein), together with a smooth deformation of the Coulomb potential in terms of a real parameter $\eta$. In a similar manner as in the Darboux III case, the Schr\"odinger equation can be analytically solved in $N$-dimensions~\cite{BEHRR2014Taub}, and therefore the information-theoretic approach to this $N$-dimensional hydrogenic system on a curved space is also feasible. This system would provide a second relevant example in which the role of the curvature could be again confronted with previous results in the literature for usual hydrogenic systems in $N$ dimensions, which are of course interesting from many different viewpoints (see, for instance~\cite{YVAS94,DAY1997entropy,Dehesa:2019nhu} and references therein). Work on all these lines is in progress and will be presented elsewhere.


\section*{Acknowledgements}

This work has been partially supported by Agencia Estatal de Investigaci\'on (Spain)  under grant  PID2019-106802GB-I00/AEI/10.13039/501100011033, by the Regional Government of Castilla y Le\'on (Junta de Castilla y Le\'on, Spain) and by the Spanish Ministry of Science and Innovation MICIN and the European Union NextGenerationEU/PRTR, as well as the contribution of the  European Cooperation in Science and Technology through the COST Action CA18108. The authors acknowledge A. Najafizade for useful discussions at the early stages of this work, and also the Referee for several relevant comments and suggestions.




\begin{landscape}

\section*{Appendix}

In the following we include Tables corresponding to the numerical data that have been used to construct some the corresponding Figures presented in the body of the paper.

\begin{table}[H]
\begin{center}
\begin{tabular}{|l||*{10}{c|}}\hline
\backslashbox{$\lambda$ }{$n$}
&\textbf{0}&\textbf{1}&\textbf{2}&\textbf{3}&\textbf{4}&\textbf{5}&\textbf{6}&\textbf{7}&\textbf{8}&\textbf{9}\\\hline\hline
\textbf{0.000} &0.5&1.5&2.5&3.5&4.5&5.5&6.5&7.5&8.5&9.5 \\\hline
\textbf{0.025} &0.494&1.445&2.349&3.207&4.022&4.795&5.529&6.224&6.884&7.508 \\\hline
\textbf{0.050} &0.488&1.392&2.207&2.941&3.6&4.192&4.722&5.198&5.623&6.005 \\\hline
\textbf{0.075} &0.482&1.341&2.075&2.7&3.231&3.681&4.063&4.386&4.662&4.896 \\\hline
\textbf{0.100} &0.476&1.292&1.952&2.483&2.91&3.252&3.527&3.75&3.931&4.078 \\\hline
\end{tabular}

\vspace{1cm}

\begin{tabular}{|l||*{10}{c|}}\hline
\backslashbox{$\lambda$ }{$n$}
&\textbf{0}&\textbf{1}&\textbf{2}&\textbf{3}&\textbf{4}&\textbf{5}&\textbf{6}&\textbf{7}&\textbf{8}&\textbf{9}\\\hline\hline
\textbf{0.000} &1&1&1&1&1&1&1&1&1&1 \\\hline
\textbf{0.025} &0.9876&0.9632&0.9395&0.9163&0.8938&0.8719&0.8506&0.8299&0.8098&0.7903 \\\hline
\textbf{0.050} &0.9753&0.9278&0.8828&0.8402&0.8&0.7621&0.7265&0.693&0.6616&0.6321 \\\hline
\textbf{0.075} &0.9632&0.8938&0.8299&0.7714&0.7179&0.6692&0.625&0.5848&0.5484&0.5154 \\\hline
\textbf{0.100} &0.9512&0.8612&0.7808&0.7095&0.6466&0.5913&0.5427&0.5&0.4624&0.4293 \\\hline
\end{tabular}

\caption{\label{table:spectrum}
Discrete spectrum $E_n^\lambda$ (above) and frequencies $\Omega_n^\lambda$ (below) for the $n=0,\ldots,9$ states and different values of $\lambda$. (Data plotted in Figure \ref{fig:spectrum}).}
\end{center}
\end{table}

\begin{table}[H]
\begin{tabular}{|l||*{16}{c|}}\hline
\backslashbox{$\lambda$ }{$n$}
&\textbf{0}&\textbf{1}&\textbf{2}&\textbf{3}&\textbf{4}&\textbf{5}&\textbf{6}&\textbf{7}&\textbf{8}&\textbf{9}&\textbf{10}&\textbf{11}&\textbf{12}&\textbf{13}&\textbf{14}&\textbf{15}\\\hline\hline
\textbf{0.000} &1.072&1.343&1.499&1.61&1.697&1.768&1.829&1.882&1.929&1.972&2.01&2.046&2.078&2.109&2.137&2.164\\\hline
\textbf{0.025} &1.091&1.374&1.539&1.658&1.751&1.828&1.894&1.952&2.003&2.049&2.091&2.129&2.164&2.196&2.226&2.254 \\\hline
\textbf{0.050} &1.109&1.403&1.577&1.701&1.799&1.879&1.947&2.006&2.058&2.104&2.145&2.184&2.219&2.251&2.282&2.311\\\hline
\textbf{0.075} &1.127&1.432&1.612&1.741&1.84&1.922&1.99&2.049&2.101&2.147&2.19&2.229&2.266&2.301&2.334&2.366 \\\hline
\textbf{0.100} &1.145&1.46&1.645&1.776&1.877&1.958&2.026&2.086&2.139&2.187&2.232&2.274&2.314&2.353&2.39&2.425\\\hline
\end{tabular}
\caption{\label{table:entropiesspace} 
Entropies $S_\rho^{n,\lambda}$ for the $n=0,\ldots, 15$ states and different values of $\lambda$. (Data plotted in Figure \ref{fig:entropiesspace}).}
\end{table}

\begin{table}[H]
\begin{tabular}{|l||*{17}{c|}}\hline
\backslashbox{$\lambda$ }{$n$}
&\textbf{0}&\textbf{1}&\textbf{2}&\textbf{3}&\textbf{4}&\textbf{5}&\textbf{6}&\textbf{7}&\textbf{8}&\textbf{9}&\textbf{10}&\textbf{11}&\textbf{12}&\textbf{13}&\textbf{14}&\textbf{15}\\\hline\hline
\textbf{0.000} &1.072&1.343&1.499&1.61&1.697&1.768&1.829&1.882&1.929&1.972&2.01&2.046&2.078&2.109&2.137&2.164 \\\hline
\textbf{0.025} &1.054&1.311&1.457&1.558&1.636&1.699&1.751&1.795&1.834&1.868&1.898&1.924&1.949&1.971&1.990&2.008 \\\hline
\textbf{0.050} &1.035&1.280&1.414&1.504&1.571&1.623&1.664&1.698&1.725&1.747&1.767&1.783&1.796&1.807&1.817&1.824 \\\hline
\textbf{0.075} &1.017&1.248&1.370&1.449&1.504&1.544&1.574&1.596&1.613&1.626&1.635&1.642&1.647&1.651&1.654&1.656  \\\hline
\textbf{0.100} &0.9998&1.217&1.327&1.394&1.437&1.466&1.486&1.499&1.508&1.513&1.517&1.519&1.520&1.520&1.520&1.519 \\\hline
\end{tabular}
\caption{\label{table:entropiesmomentum} 
Entropies $S_\gamma^{n,\lambda}$ for the states with $n=0,\ldots, 15$ and different values of $\lambda$. (Data plotted in Figure \ref{fig:entropiesmomentum}).}
\end{table}

\begin{table}[H]
\begin{tabular}{|l||*{17}{c|}}\hline
\backslashbox{$\lambda$ }{$n$}
&\textbf{0}&\textbf{1}&\textbf{2}&\textbf{3}&\textbf{4}&\textbf{5}&\textbf{6}&\textbf{7}&\textbf{8}&\textbf{9}&\textbf{10}&\textbf{11}&\textbf{12}&\textbf{13}&\textbf{14}&\textbf{15}\\\hline\hline
\textbf{0.000} &2.145&2.685&2.997&3.219&3.393&3.536&3.658&3.764&3.858&3.943&4.020&4.091&4.156&4.217&4.274&4.327 \\\hline
\textbf{0.025} &2.145&2.685&2.995&3.216&3.387&3.527&3.645&3.747&3.837&3.917&3.988&4.053&4.113&4.167&4.217&4.262 \\\hline
\textbf{0.050} &2.145&2.683&2.990&3.206&3.370&3.502&3.611&3.703&3.783&3.851&3.913&3.966&4.015&4.058&4.099&4.135 \\\hline
\textbf{0.075} &2.145&2.681&2.982&3.190&3.345&3.466&3.564&3.645&3.714&3.773&3.825&3.871&3.913&3.952&3.988&4.021  \\\hline
\textbf{0.100} &2.145&2.677&2.972&3.170&3.314&3.424&3.512&3.585&3.646&3.700&3.749&3.793&3.834&3.872&3.909&3.944 \\\hline
\end{tabular}
\caption{\label{table:entropiessum} 
$S_\rho^{n,\lambda} + S_\gamma^{n,\lambda}$ for the $n=0,\ldots, 15$ states and different values of $\lambda$. (Data plotted in Figure \ref{fig:entropiessum}).}
\end{table}

\begin{table}[H]
\begin{center}
\begin{tabular}{|l||*{3}{c|}}\hline
\backslashbox{$\lambda$ }{$n$}
&\textbf{0}&\textbf{1}&\textbf{2}\\\hline\hline
\textbf{0.00} &2.145&2.685&2.997 \\\hline
\textbf{0.25} &2.146&2.648&2.887 \\\hline
\textbf{0.50} &2.154&2.604&2.775 \\\hline
\textbf{0.75} &2.173&2.581&2.718 \\\hline
\textbf{1.00} &2.199&2.570&2.689 \\\hline
\end{tabular}
\caption{ \label{table:entropiessumlambdabig} 
$S_\rho^{n,\lambda} + S_\gamma^{n,\lambda}$ for the $n=0,1,2$ states and large values of $\lambda$. (Data plotted in Figure \ref{fig:entropiessumlambdabig}).}
\end{center}
\end{table}

\begin{table}[H]
\begin{center}
\begin{tabular}{|l||*{10}{c|}}\hline
\backslashbox{$\lambda$ }{$n$}
&\textbf{0}&\textbf{1}&\textbf{2}&\textbf{3}&\textbf{4}&\textbf{5}&\textbf{6}&\textbf{7}&\textbf{8}&\textbf{9}\\\hline\hline
\textbf{0.00} &3.217&4.151&4.709&5.109&5.421&5.676&5.893&6.082&6.248&6.397 \\\hline
\textbf{0.01} &3.255&4.235&4.839&5.284&5.639&5.938&6.198&6.428&6.636&6.826 \\\hline
\textbf{0.02} &3.292&4.319&4.967&5.454&5.850&6.188&6.485&6.750&6.992&7.214 \\\hline
\textbf{0.03} &3.329&4.402&5.093&5.619&6.051&6.422&6.749&7.043&7.309&7.553 \\\hline
\textbf{0.04} &3.366&4.484&5.215&5.777&6.241&6.639&6.990&7.303&7.587&7.845 \\\hline
\end{tabular}
\caption{\label{table:entropiesspace3d}
Entropies $S_\rho^{3,n,0,0, \lambda}$ for the states with $n=0,\ldots, 9$ and different values of $\lambda$. (Data plotted in Figure \ref{fig:entropiesspace3d}).}
\end{center}
\end{table}

\begin{table}[H]
\begin{center}
\begin{tabular}{|l||*{10}{c|}}\hline
\backslashbox{$\lambda$ }{$n$}
&\textbf{0}&\textbf{1}&\textbf{2}&\textbf{3}&\textbf{4}&\textbf{5}&\textbf{6}&\textbf{7}&\textbf{8}&\textbf{9}\\\hline\hline
\textbf{0.00} &3.217&4.151&4.709&5.109&5.421&5.676&5.893&6.082&6.248&6.397 \\\hline
\textbf{0.01} &3.180&4.066&4.577&4.931&5.198&5.408&5.580&5.724&5.846&5.950 \\\hline
\textbf{0.02} &3.142&3.980&4.445&4.753&4.974&5.140&5.268&5.369&5.449&5.514 \\\hline
\textbf{0.03} &3.105&3.895&4.313&4.576&4.753&4.878&4.967&5.031&5.077&5.111 \\\hline
\textbf{0.04} &3.068&3.810&4.182&4.401&4.538&4.626&4.683&4.718&4.740&4.752 \\\hline
\end{tabular}
\caption{\label{table:entropiesmomentum3D}
Entropies $S_\gamma^{3,n,0,0, \lambda}$ for the states with $n=0,\ldots, 9$ and different values of $\lambda$. (Data plotted in Figure \ref{fig:entropiesmomentum3D}).}
\end{center}
\end{table}

\begin{table}[H]
\begin{center}
\begin{tabular}{|l||*{10}{c|}}\hline
\backslashbox{$\lambda$ }{$n$}
&\textbf{0}&\textbf{1}&\textbf{2}&\textbf{3}&\textbf{4}&\textbf{5}&\textbf{6}&\textbf{7}&\textbf{8}&\textbf{9}\\\hline\hline
\textbf{0.00} &6.434&8.301&9.418&10.218&10.841&11.353&11.786&12.163&12.496&12.794 \\\hline
\textbf{0.01} &6.434&8.301&9.417&10.215&10.837&11.346&11.778&12.151&12.481&12.776 \\\hline
\textbf{0.02} &6.434&8.300&9.413&10.207&10.824&11.328&11.753&12.119&12.441&12.728 \\\hline
\textbf{0.03} &6.434&8.297&9.406&10.195&10.804&11.300&11.716&12.073&12.386&12.664 \\\hline
\textbf{0.04} &6.434&8.294&9.397&10.178&10.779&11.265&11.672&12.021&12.326&12.598 \\\hline
\end{tabular}
\caption{\label{table:entropiessum3d}
Entropies $S_{\mathrm{total}}^{3,n,0,0, \lambda}$ for the states with $n=0,1,\dots,9$ and different values of the curvature $\lambda$. (Data plotted in Figure \ref{fig:entropiessum3d}).}
\end{center}
\end{table}

\end{landscape}



\end{document}